\documentclass[twocolumn]{aastex631}

\newcommand\suzaku{{\it Suzaku}}
\newcommand\xmm{{\it XMM--Newton}}

\newcommand\astrosat{{\it AstroSat}}
\newcommand\nustar{{\it NuSTAR}}

\newcommand\swift{{\it Swift}}
\usepackage{tabularx}
\usepackage{multirow}
\usepackage{amsmath}

\begin{document}

%\linenumbers

\title{Exploring the Accretion disc/Corona Connection in NGC~6814: Insights from UV and X--ray spectral--timing studies}

\author{Kavita Kumari}
\affiliation{Inter-University Center for Astronomy and Astrophysics, Pune 411007, India}
\affiliation{Department of Physics, Kamla Rai College, Constituent unit of Jai Prakash University, Gopalganj, Bihar 841428, India}

\author{I. E. Papadakis}
\affiliation{Department of Physics and Institute of Theoretical and Computational Physics, University of Crete, 71003 Heraklion, Greece}
\affiliation{Institute of Astrophysics, FORTH, GR-71110 Heraklion, Greece}

\author{G.C. Dewangan}
\affiliation{Inter-University Center for Astronomy and Astrophysics, Pune 411007, India}

\begin{abstract}

We conducted a comprehensive spectral and timing analysis of NGC~6814 using \astrosat{}'s 2019 and \xmm{}'s 2021 observations.
Cross--correlation analysis revealed a significant correlation between FUV (1541 \AA)/X--ray and UVW1 (2910 \AA)/X--ray variations, with delays of $\sim 15$~ks and 30~ks, respectively. We constructed four broadband SEDs after applying aperture correction (for the UVIT filter), subtracting host galaxy and emission line contributions from UV flux, and using mean X--ray spectra alongside selected UV data points. 
First, we fitted the SEDs with  \textsc{kynsed} model assuming various combinations of inclination, $\theta$, color correction factors, $f_{\rm col}$, and BH spins. Best--fit models were achieved for $\theta=70^{\circ}$ (consistent with past estimates for this source) and for spin $\leq 0.5$, 
while $f_{\rm col}$ is not constrained. 
\textsc{kynsed} provided satisfactory fit to all SEDs in the case when the corona is powered by the accretion process, with $\sim 10-20$\% of the accretion power transferred to the corona,  $\dot{m}/\dot{m}_{\rm Edd}\sim 0.1$, corona radius of $\sim 6-10~r_g$, and height of $\sim7.5-35~r_g$.
Model time--lags computed using the SED best--fit results are aligned well with the observed time--lags. 
Although some of the model parameters are not constrained, the important result of our work is that both the broadband X--ray/UV spectra and the X--ray/UV time--lags in NGC~6814 are consistent with the hypothesis of X--ray illumination of the disc in a lamp--post geometry framework. Within this model framework, we do not need to assume an outer or inner truncated disc.

\end{abstract}

 
\section{Introduction}\label{sec:intro}

The study of the inner region of the Active Galactic Nucleus (AGN) is essential for unravelling the physics of black hole accretion. Understanding the geometry of the X--ray corona, the process of accretion and the interplay between these two in the inner region of AGN are well-known challenges due to limitations in the spatial resolution of current telescopes (in all bands).
However, correlation studies between the X--ray and optical/Ultraviolet (UV) continuum light curves, and broadband spectral analysis provide in-direct methods to probe this region. 

In principle, if the X--ray and UV/optical emission processes are coupled, the study of correlations (and time--lags) between X–rays and UV/optical variations in AGN can provide essential clues on their connections and provide information about the disc-corona geometry. This possibility has been explored in numerous studies in past decades (see e.g. \citealt{Nandra_1998ApJ, Breedt_2009MNRAS,  Troyer_ngc6814_2016MNRAS, mchardy_ngc4593_2018MNRAS, edelson_ngc4151_2017ApJ, edelson_2019ApJ, Cackett_2020ApJ, Hernandez_2020mnras, Kara_AGNSTORM_2021ApJ, Miller_mrk876_2023ApJ, Kumari2023MNRAS, Kavita_ngc4051_2024MNRAS}). 

Recently, a detailed theoretical study of the expected correlations/time--lags in the case of the X--ray lamp--post geometry has been put forward by \cite{ Kammoun_2021ApJ}. They computed the disc response function ($\Psi$) assuming a standard Novikov–Thorne (NT) accretion disc (\citealt{NT1973}), taking into consideration all special and general relativistic effects as well as the disc ionization effects. They have demonstrated that, if all the effects are treated properly, the theoretical model on thermal disc reprocessing matches the observed lag spectra very well \citep{Kammoun_2019ApJ,  Kammoun_2021MNRAS, kammoun_lag_2023MNRAS}. This has been observationally explored for NGC~4051 in the work by \cite{Kavita_ngc4051_2024MNRAS}.

Furthermore, a new model, \textsc{kynsed}, has been developed by \cite{Dovciak_kynsed2022} to fit the broadband spectral energy distribution (SED) of AGN. This model is like other models, for example \textsc{optxagnf} \citep{Done_2012MNRAS} and \textsc{agnsed} \citep{Kubota_agnsed_2018MNRAS}, which can also fit broadband SEDs from X--rays to UV/optical in AGN. However, contrary to these models, \textsc{kynsed} can compute the SED in the case when the accretion disc is illuminated by X--rays, in the lamp--post geometry. The X--ray corona can be powered by an unknown source but also by the accretion process itself. In the latter case, the accretion power below a transition radius ($R_{t}$) is transferred to the corona through some unknown physical mechanism. This assumption establishes a direct link between the accretion disc and the X--ray source.  
The model also considers the effect of electron scattering in the disc atmosphere by introducing various color correction factors ($f_{\rm col}$). In this work, we aim to investigate if  \textsc{kynsed} can fit the broadband SED in another nearby AGN, NGC~6814. We also make the connection between the results obtained using cross--correlation studies and spectral fitting.

NGC~6814 is a nearby ($z=0.00522$) highly variable Seyfert type 1.5 AGN. It was a part of the Lick AGN Monitoring Project (LAMP; \citealt{bentz_LAMP_2009ApJ}).  Its black hole (BH) mass estimate from the reverberation mapping technique is $M_{\rm BH}= 1.09^{+0.15}_{-0.14}\times 10^7 M_{\odot}$ \citep{bentz_massdatabase2015}.  Previously, \cite{Pancoast_2014MNRAS} performed dynamical modelling using the LAMP data and found a lower BH mass of $M_{\rm BH} =
2.6^{+ 1.9}_{- 0.9} \times 10^6 M_{\odot}$ and broad-line region (BLR) inclination $i = 49^{+20\circ}_{-22}$. The Cepheid-based distance of NGC~6814 is $21.65\pm 0.41$~Mpc \citep{Bentz_2019ApJ_cepheid_distance}. 

\suzaku{} observations during 2011 showed strong X--ray variability, and iron K$\alpha$~(6.4 keV) \& K$\beta$~(6.94 keV) lines but no evidence of soft X--ray excess emission \citep{Walton_ngc6814_2013ApJ}. 
\cite{Malizia_2014ApJ} conducted a broadband spectral analysis using simultaneous data from \xmm{}, {\it INTEGRAL/IBIS} and {\it Swift/BAT}, reporting a high--energy cut--off for NGC~6814 at $E_{\rm cut} = 190^{+185}_{-66}$~keV. Later, \cite{Tortosa_2018_census}, using \nustar{} observations, determined the cut--off to be $E_{\rm cut} = 135^{+70}_{-35}$~keV. \cite{Molina_2019MNRAS}, analyzing data from {\it Swift/XRT}  and \nustar{}, found that the broadband X--ray emission from NGC~6814 is primarily characterized by a relatively flat power-law with a photon index of $\Gamma \sim 1.68$ and a high--energy cut--off of $E_{\rm cut} = 115^{+26}_{-18}$~keV.
Recently, \cite{Gallo_ngc6814_2021ApJ} have reported the detection of a rapid occultation event in 2016 observation by \xmm{}. They computed the properties of the obscuring gas and claimed that it must be located in the BLR. They also reported a high disc inclination ($67^{+1\circ}_{-4}$). \cite{Kang_2023MNRAS} further explored the eclipsing event with the hardness ratio (HR)--count rate (CR) plot and suggested that the eclipsing absorber might be clumpy instead of a single cloud. They also investigated the 2009 (former) and 2021 (later) observations by \xmm{} and found that a warm absorber was present only in the 2016 observation. 

Using over 3 months of monitoring data observed by \swift{} and \textit{Liverpool} Telescope in 2012,  \cite{Troyer_ngc6814_2016MNRAS} found that the variations in UV and optical bands lag behind the X--ray by $\sim2$ days and support the scenario of thermal disc reprocessing. Recently, \cite{Gonzalez_2024MNRAS_ngc6814} utilized the observations from the 2022 \swift{} monitoring campaign over 75 days and found that the UV/optical and X--ray variability is most likely driven by separate mechanisms. The cross--correlation analysis reveals that the X--rays lead the UV ($\rm 1928$~\AA) by $\sim0.4$~days while variations in optical bands lag UV by a constant lag of $\sim0.3$~days. 
They found that no combination of physical parameters can simultaneously explain the observed time--lag spectrum and AGN SED in a self--consistent way. Based on their findings, they hint at the possibility of a non--standard geometry in NGC~6814.

In this work, we are using the \astrosat{} and \xmm{} observations for spectral and timing studies using the \textsc{kynsed} model. We have explained the data reduction process in section \ref{sec:data} followed by the timing and spectral analysis in section \ref{sec:timing_analysis} and \ref{sec:spec_analysis}, respectively. In Section \ref{sec:spectral_timing} we have tried to build a connection between spectral and timing results. We have discussed our results in section \ref{sec:discussion}.

%%%%%%%%%%%%%%%%%%%%%%%%%%%%%%%%%%%%%%%%%%%%%%%%%%%%%%%%%%%%%%%%%%%%%%%%%%%%%%%%%
%%%%%%%-------- section2: Observation and Data reduction ------------%%%%%%%%%%%%
%%%%%%%%%%%%%%%%%%%%%%%%%%%%%%%%%%%%%%%%%%%%%%%%%%%%%%%%%%%%%%%%%%%%%%%%%%%%%%%%%

\section{Observations and Data reduction} \label{sec:data}

%---------------Figure 1: Light curves -------------------
\begin{figure*}[!ht]
\centering
\includegraphics[width=0.6\textwidth]{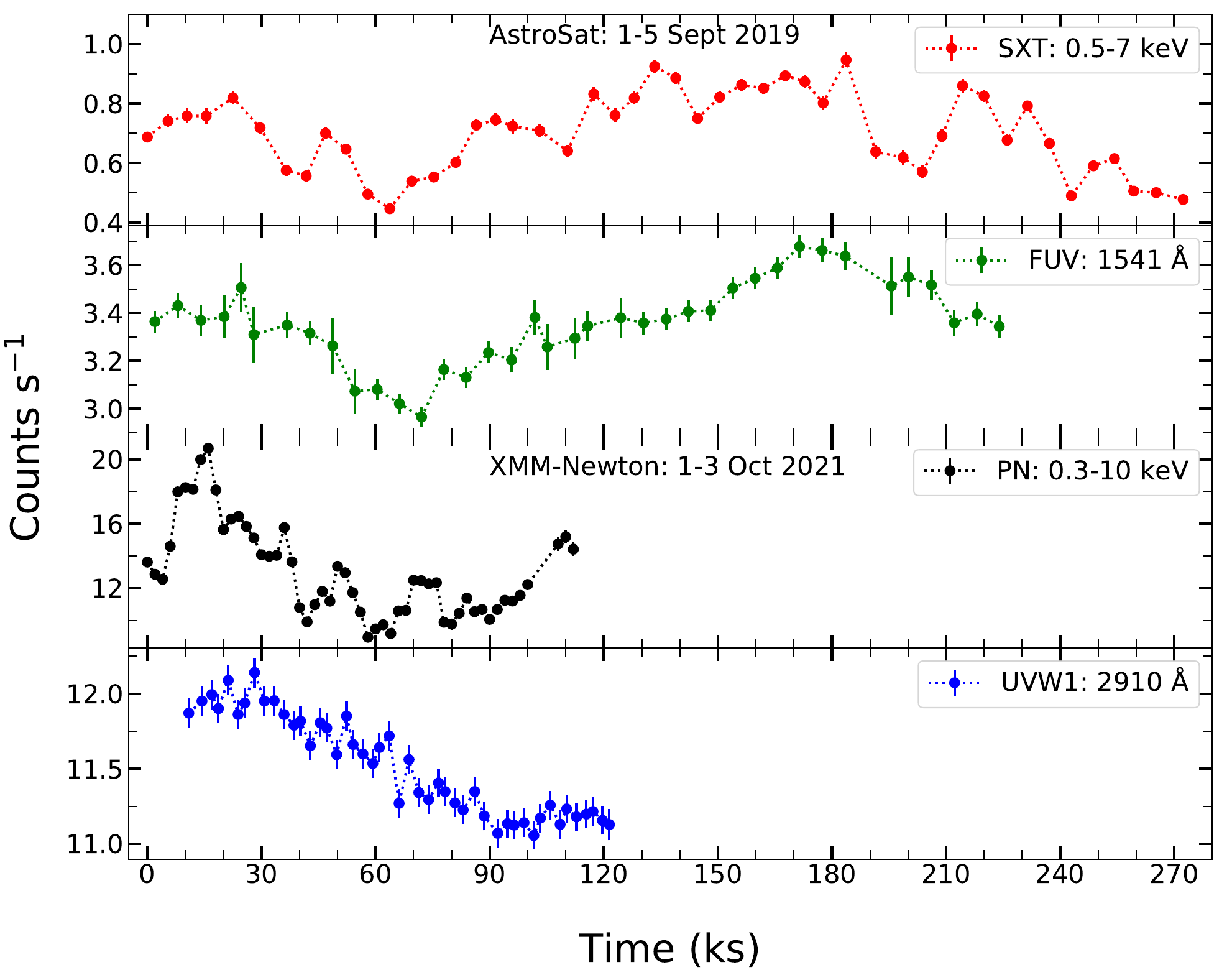}
\caption{Light curves of NGC~6814 extracted from   \astrosat{} and the 2021 \xmm{} observations. The SXT (0.5--7 keV) and UVIT/FUV ($\lambda_{\rm mean} =1541 $~\AA) light curves are shown in red and green, respectively. The EPIC--PN (0.3--10 keV) and OM/UVW1 ($\lambda_{\rm mean} = 2910 $~\AA) light curves are plotted in black and blue, respectively.}
\label{fig:lc}
\end{figure*}

%%%%%%%%%%%%%%%%%%%%%%%%%%%%%%%%%%%%%%%%%%%%%%%%%%%%%%%%%%%%%
%%%%%%%--------------------- TABLE 1 data information --------
%%%%%%%%%%%%%%%%%%%%%%%%%%%%%%%%%%%%%%%%%%%%%%%%%%%%%%%%%%%%%

\begin{deluxetable*}{ccccccc}
\tablecaption{Details of the \astrosat{} and \xmm{} observations and fractional variability amplitude ($F_{var}$).}\label{tab:observation}
\tablehead{
\colhead{Obs. Date} & \colhead{Satellite} & \colhead{Obs. ID} & \colhead{Energy band/Filter} & \colhead{Exp. time} & \colhead{Mean counts}  & \colhead{$F_{\rm var}$} \\
\colhead{(dd/mm/yyyy}) & \colhead{} & \colhead{}  & \colhead{} & \colhead{(ks)} &\colhead{($\rm counts~s^{-1}$)}&  \colhead{(\%)}
}
\colnumbers
\startdata 
1--5 Sept 2019 & \astrosat{} & A05\_037T01\_9000003140 & 0.5--7 keV & 91.1 & 0.71  & $18.68 \pm 0.41$\\
 & & &  F154W: 1541 \AA & 46.7 & 3.36  & $4.69 \pm 0.34$\\
1--3 Oct 2021 & \xmm{} & 0885090101 & 0.3--10 keV & 80.9 & 13.03  & $22.1 \pm 0.16$ \\
 & & &  UVW1: 2910 \AA & 94 & 11.52   &  $2.76 \pm 0.13$  \\
\enddata
\end{deluxetable*}

\subsection{AstroSat}\label{subsec:astrosat_data}

\astrosat{} observed NGC~6814 during 1--5 September 2019 (Obs.~ID:~A05\_037T01\_9000003140) with Soft X--ray Telescope (SXT; \citealt{singh_2016SPIE, singh_2017JApA}) as the primary instrument. The source was also observed simultaneously with the Ultra--Violet Imaging Telescope (UVIT; \citealt{Tandon_2017AJ, Tandon_2020AJ}) in the FUV band with F154W/BaF2 filter ($\lambda_{\rm mean}=1541$~\AA).
We downloaded the Level1 (L1) UVIT and orbit--wise SXT data from the \astrosat{} archive\footnote{\url{https://astrobrowse.issdc.gov.in/astro_archive/archive/Home.jsp}}. 

%%%%%%%%%%%%%%%%%%%%%%%%%%%%%%%%%%%%%%%%%%%%%%%%%%%%%
%%%----------AstroSat SXT data reduction------------
%%%%%%%%%%%%%%%%%%%%%%%%%%%%%%%%%%%%%%%%%%%%%%%%%%%%%

We processed the SXT data using the {\sc sxtpipeline}\footnote{\url{https://www.tifr.res.in/~astrosat_sxt/index.html}} (Version:~1.4b) to get the orbit--wise clean event file. We then merged them using the Julia package  {\textsc SXTMerger.jl}\footnote{\url{https://github.com/gulabd/SXTMerger.jl}} to create a merged event file which can be used to extract the images and light curves using the heasoft tool \textsc{xselect}. We used the circular region of radius 15~arcmin to extract the source light curve in the 0.5--7.0~keV band with the bin size of 2.3775~sec (time resolution of SXT). We then rebinned the light curve with the bin size of 5820~sec (97 minutes), the orbital time of the \astrosat{}. The reasoning and technique of rebinning have been discussed in \cite{Kumari2023MNRAS}. 

%%%%%%%%%%%%%%%%%%%%%%%%%%%%%%%%%%%%%%%%%%%%%%%%%%%%%
%%%----------AstroSat UVIT data reduction------------
%%%%%%%%%%%%%%%%%%%%%%%%%%%%%%%%%%%%%%%%%%%%%%%%%%%%%

The simultaneous L1 UVIT data was processed using the \textsc{ccdlab} UVIT Pipeline \citep{Postma_2017PASP}. The pipeline extracts and applies all the necessary corrections on the L1 UVIT data (such as centroiding bias, flat-fielding etc.). We also corrected the pointing drift in each orbit using the VIS images taken every second and generated orbit--wise cleaned images and centroid lists. We then aligned all the orbit--wise images to create a merged image and centroid list.  The radii of circular regions for the source and background counts are 16~pixels or 6.66~arcsec (1 pixel = 0.416~arcsec) and 80~pixels (33.28~arcsec), respectively.  We used the Julia--based {\sc uvitools} package (described in \citealt{dewangan_2021JApA}) to extract background--subtracted source counts from the orbit--wise images. We have shown the extracted SXT (in red) and FUV (in green) light curves in Fig.~\ref{fig:lc}. We also used the merged image to calculate the growth curve of the point spread function (PSF) to perform aperture correction and measure the host galaxy's contribution to the flux detected within the extraction region.
For details, see Appendix \ref{appendix1} and \ref{appendix2}.

%%%%%%%%%%%%%%%%%%%%%%%%%%%%%%%%%%%%%%%%%%%%%%%%%%%%%
%%%----------XMM-Newton data reduction------------
%%%%%%%%%%%%%%%%%%%%%%%%%%%%%%%%%%%%%%%%%%%%%%%%%%%%%

\subsection{XMM--Newton}\label{subsec:xmm_data}

NGC~6814 was observed by the \xmm{}~(\citealt{Jansen_xmm_2001}) 
during  22 April 2009 (Obs.~ID:~	
0550451801; XMM09), 8--9 April 2016 (Obs.~ID:~0764620101; XMM16)  and
1--3 October 2021 (Obs.~ID:~0885090101; XMM21). The XMM09 observation is very short, with an exposure of $\sim 32$ ks, which is approximately 1/4 of the other two XMM-Newton observations.  Moreover, the Optical Monitor
(OM) was operated in imaging mode using the U--band filter only ($\lambda_{\rm mean}$ = 3440~\AA), resulting in just 10 data points. Both the short exposure and the very few points in the U--band light curve suggest that this observation is not appropriate for a cross-–correlation
analysis.

%---------------Figure 2: XMM16 Light curves -------------------
\begin{figure}
\centering
\includegraphics[width=0.46\textwidth]{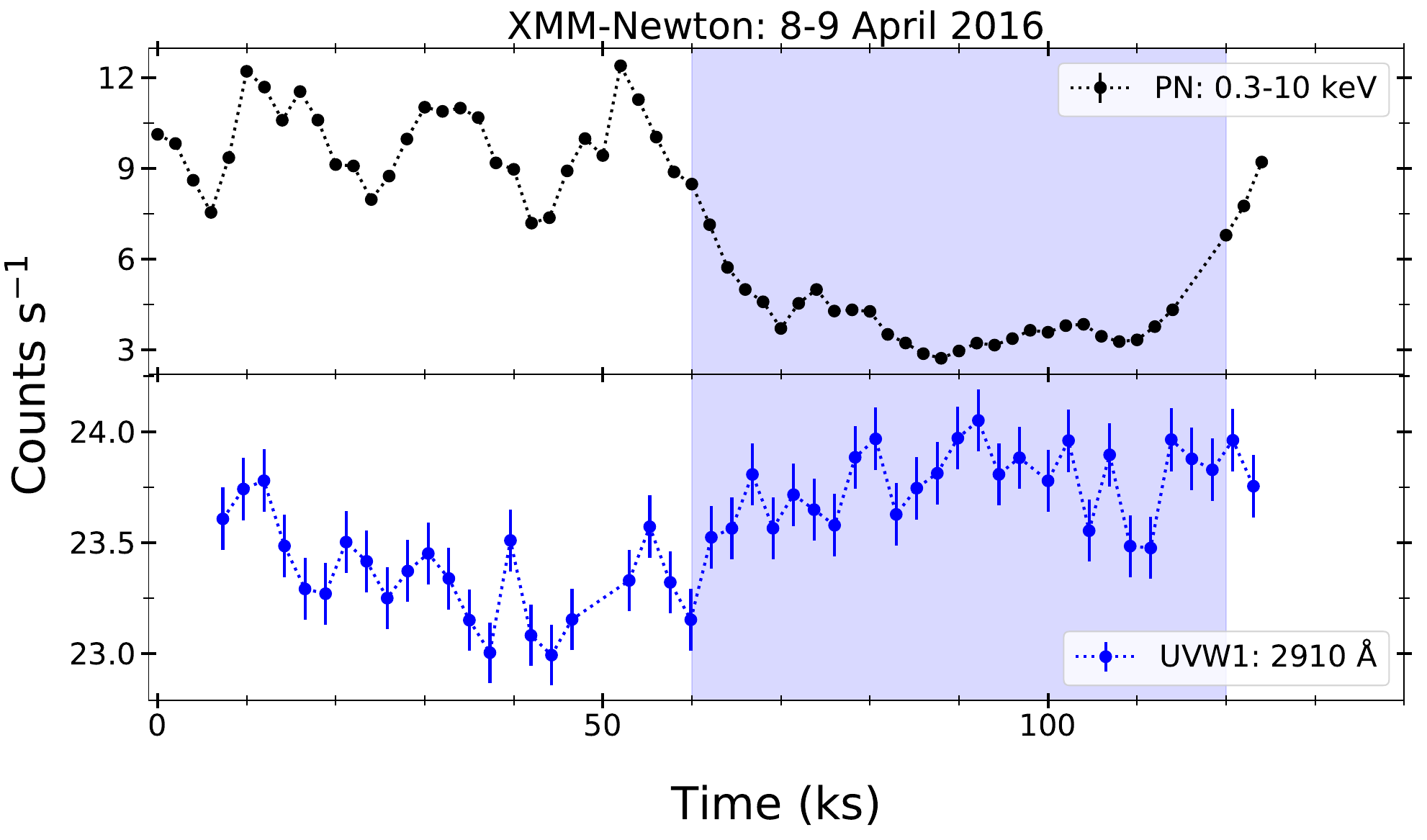}
\caption{EPIC--PN (0.3--10 keV) and OM/UVW1 light curves of NGC~6814 extracted from  the 2016 \xmm{} observation.  The blue shaded part indicates the X--ray band occultation event reported by \cite{Gallo_ngc6814_2021ApJ} and \cite{Kang_2023MNRAS}.}
\label{fig:lcxmm16}
\end{figure}

\subsubsection{The X--ray eclipses} The XMM16 observation is known for the X--ray occultation event that occurred $\sim 60$~ ks after the start of the observation \citep{Gallo_ngc6814_2021ApJ}. Figure \ref{fig:lcxmm16} shows the EPIC--PN and the OM/UVW1 light curve for this observation. The blue shaded parts in this figure indicate the time period of the X--ray occultation event in this source, as reported and studied by \cite{Gallo_ngc6814_2021ApJ} and \cite{Kang_2023MNRAS}. In fact, 
\cite{Kang_2023MNRAS} claims that the corona was already eclipsed at the beginning of the XMM16 exposure. Figure \ref{fig:lcxmm16} clearly shows that the UV light curve does not appear to be significantly affected by this event. This could be because the diameter of the obscurer is probably much smaller than the size of the accretion disc, which emits UV light in this source. Given that the X--ray and UV light curves are clearly affected by external, different variability processes, we cannot use the XMM16 data for our analysis.

\begin{figure}
\centering
\includegraphics[width=0.46\textwidth]{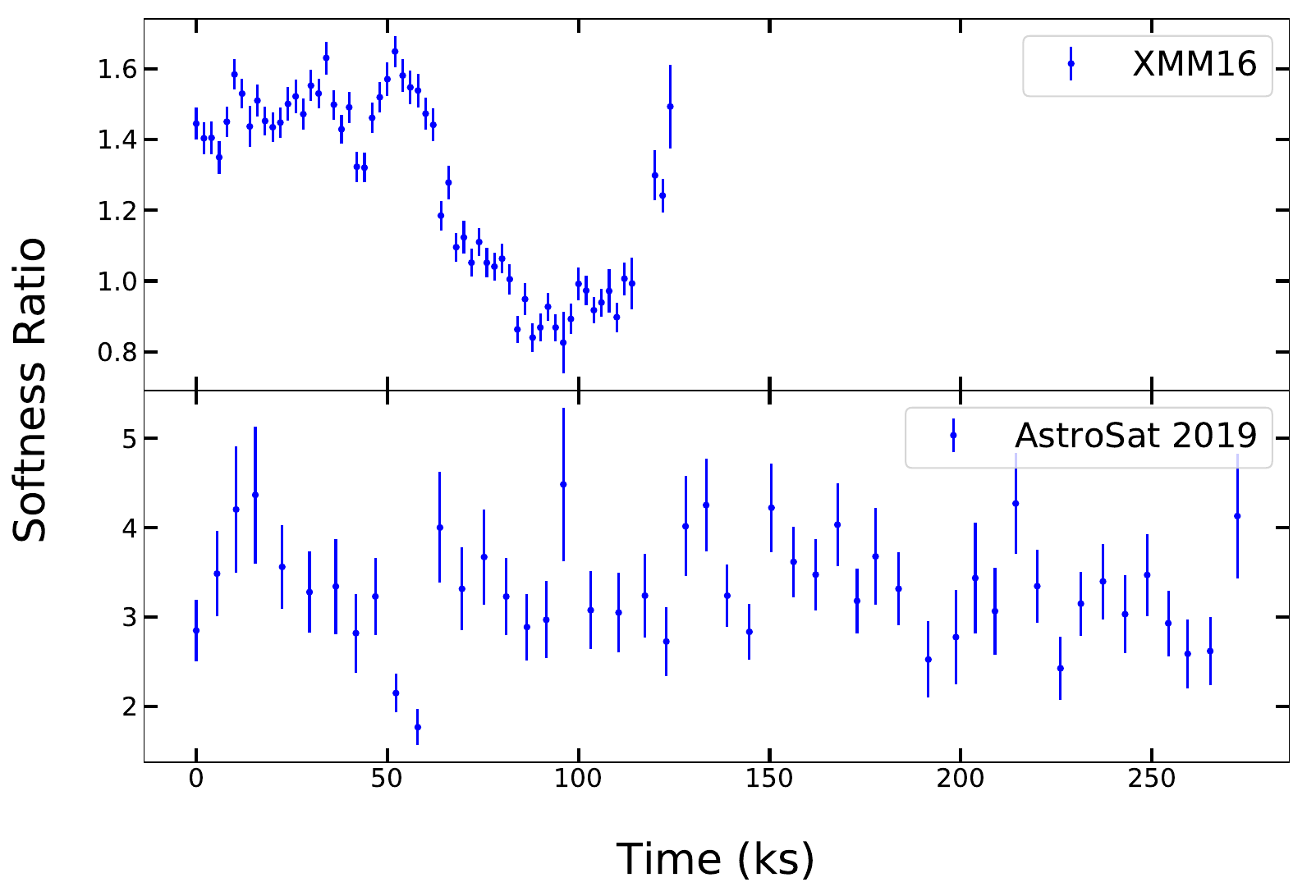}
\caption{The upper and lower panels show the softness ratio (2--4 keV/4--7 keV) for XMM16 and the \astrosat{} observations, respectively.}
\label{fig:softness_ratio}
\end{figure}

\cite{Kang_2023MNRAS} showed that the XMM21 observation is not affected by X--ray eclipsing events. We searched the {\it AstroSat} light curve for such events, by creating 2--4 and 4--7 keV band light curves and then computing the (2--4 keV)/(4--7 keV) ratio or Softness ratio. The upper panel of Fig. \ref{fig:softness_ratio} shows the softness ratio for the XMM16 observation.  This plot shows the fast, and large amplitude evolution of this ratio during the eclipsing event in the second part of the XMM16 observation. The bottom panel of the same figure shows the evolution of the softness ratio for the \astrosat{} observation. There are no such fast--evolving and large amplitude variations in the softness ratio plot for the \astrosat{} observation. We therefore conclude that the \astrosat{} observation is not affected by such events.

%--------Figure 3: Softness ratio -------------------

\subsubsection{The {\it XMM-Newton} data analysis} 
For XMM21, the European Photon Imaging Camera (EPIC--PN; \citealt{Struder_epicpn_2001}) was operated in the Large Window mode with the medium filter and the Optical/UV Monitor telescope (OM; \citealt {Mason_xmm_om_2001}) was operated only in the \textit{imaging} mode. The majority of these images were obtained using the UVW1 filter, which has a central wavelength of 2910~\AA. Observations in other bands (U, V, B, UVM2, and UVW2) consist of only a single exposure at the beginning of the OM observation. Consequently, we only utilized the observations in the UVW1 band for our analysis.
We downloaded the Observation Data Files (ODFs) from the \xmm{} science archive\footnote{\url{http://nxsa.esac.esa.int/nxsa-web/\#search}} (XSA). We reduced the data acquired with the EPIC--PN and OM using the Science Analysis System (SAS v20.0.0) software and the latest calibration files. 

%%%%%%%%%%%%%%%%%%%%%%%%%%%%%%%%%%%%%%%%%%%%%%%%%%%%%
%%%----------OM data reduction------------
%%%%%%%%%%%%%%%%%%%%%%%%%%%%%%%%%%%%%%%%%%%%%%%%%%%%%

We extracted the OM images from the ODFs using the SAS image processing chain {\sc omichain}. 
This SAS processing chain performs all the necessary corrections to the data and provides fully processed OM images with the observed source list for each exposure. The pipeline performs standard aperture photometry to obtain the background subtracted count rates, applies Astrometric corrections to each source's position and converts it to the sky coordinate. We have run the task {\sc omichain} with the default input parameters.
The source extraction region is 6~pixels radius or 5.7 arcsec (1 pixel = 0.953 arcsec). We obtained 47 data points in the light curve from XMM21 observations with a mean time gap of  $\sim 2.4 $~ks between the data points. We show the final extracted UVW1 light curves in the last panel (in blue) in Fig. \ref{fig:lc}.

%%%%%%%%%%%%%%%%%%%%%%%%%%%%%%%%%%%%%%%%%%%%%%%%%%%%%
%%%----------EPIC-PN data reduction------------
%%%%%%%%%%%%%%%%%%%%%%%%%%%%%%%%%%%%%%%%%%%%%%%%%%%%%

We processed the ODFs to generate the calibrated and concatenated EPIC--PN event lists using {\sc epproc} task. We extracted the count rate in 10--12 keV energy band to look for the presence of a high--energy flaring particle background. We excluded the flaring periods by defining low background intervals with {\sc rate<=0.35} and generated a good time interval (GTI) file using the {\sc tabgtigen} task for both observations. Then we created a filtered event file using the GTI file which we used to extract the light curves and spectrum.
We also checked for pile--up and found that our data was not affected by it. We choose a circular source extraction region of radius 35~arcsec centred at the source position and a background region of the same radius from the source-free region on the same detector chip. We used the task {\sc evselect} to extract the source and background light curves from the selected regions with a bin size of 2~ks and obtained the background subtracted light curve using task {\sc epiclccorr}.
These light curves are shown in black in Fig.~\ref{fig:lc}. Other observational details are tabulated in Table \ref{tab:observation}.

%%%%%%%%%%%%%-----------------Figure 4  ------------%%%%%%%%%%%%%%%%%%%
\begin{figure}[!ht]
\centering
\includegraphics[width=1\columnwidth]{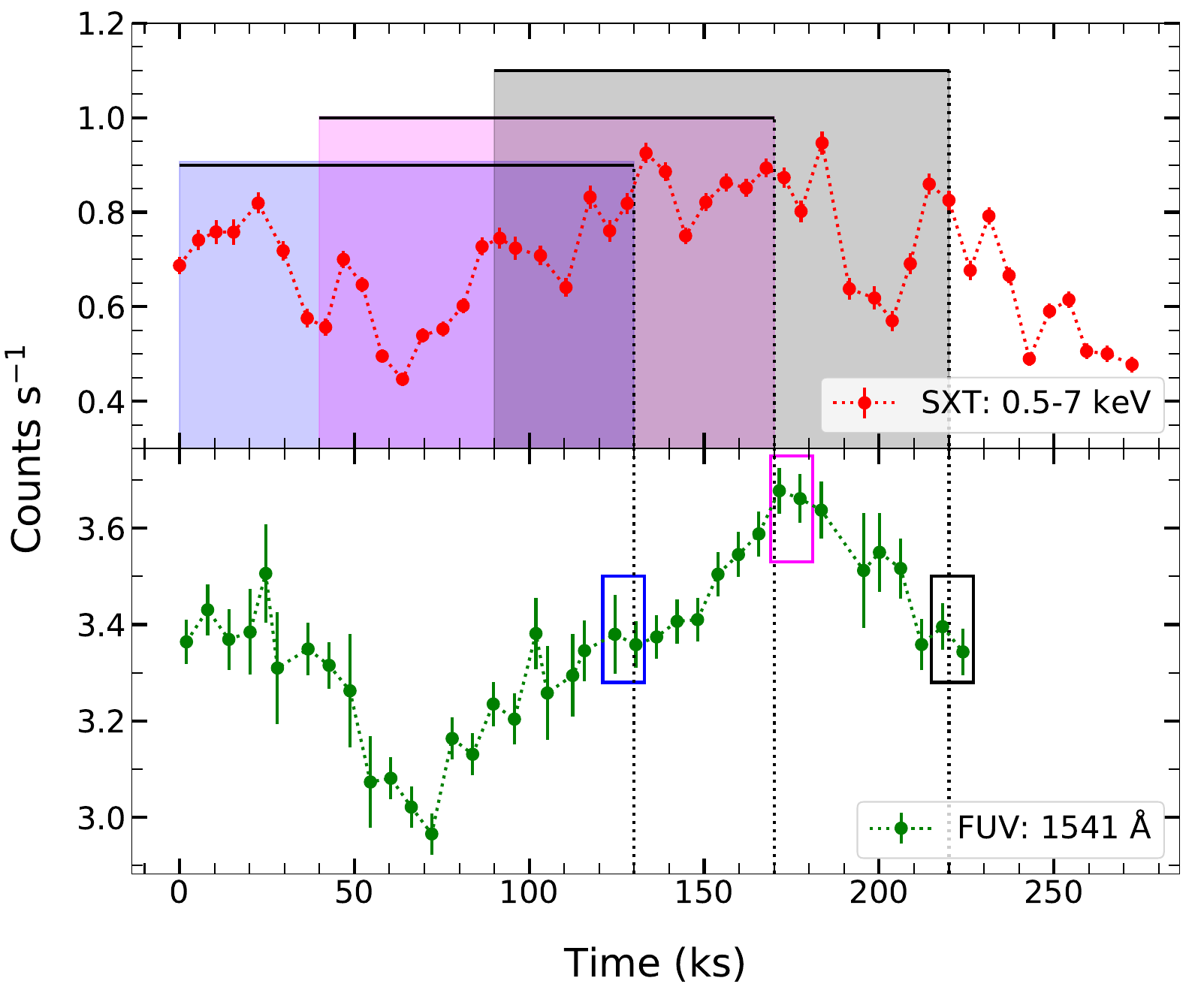}
\caption{ The observed SXT and FUV light curves from \astrosat{} 2019 observation have been plotted in red and green, respectively. The shaded regions in the upper panel correspond to the time intervals over which we integrate to extract the mean X--ray spectrum. We extracted three spectra between 0--130 ks (blue), 40--170 ks (magenta) and 90--220 ks (grey) regions. The vertical dotted lines indicate the times at which we select the UV data points in the lower panel. We have marked the corresponding FUV data points with a rectangular box, (see text in section \ref{sec:SED} for details). }
\label{fig:astrosat_lc}
\end{figure}

\subsection{Creating time resolved SEDs} \label{sec:SED}

One of the main objectives of our work is to investigate whether the X--ray reverberation hypothesis can explain both the time delays and the broadband X--ray to UV spectral variability in this source. If indeed the observed variability in the UV/optical bands is due to X--ray reverberation, then the time-variable flux at any wavelength, $\lambda$, can be written as,

\begin{equation}
F_{\rm obs} (\lambda,t)=F_{\rm NT}(\lambda)+\int_{0}^{+\infty} L_{\rm X}(t-t')\Psi_{L_X}(\lambda,t')dt',
\label{eq:eq1}
\end{equation}

\noindent where, $F_{\rm NT}(\lambda)$ is the emission of the accretion disc in the $\lambda$ band, $L_{\rm X}(t)$ is the variable X--ray luminosity, and $\Psi_{L_X}(\lambda,t)$ is the response function of the accretion disc in the spectral band with centroid wavelength $\lambda$. The equation above shows that the variable disc emission in each epoch depends on the past X--ray flux and on the disc response function \citep[for details see e.g. \S~2 in ][]{Dovciak_kynsed2022}. The latter determines the time period over which the X--ray luminosity affects the current emission of the accretion disc in the UV/optical bands.

Equation \ref{eq:eq1} shows that, under the hypothesis of X--ray reverberation, if we want to study the variable broadband SEDs, we cannot construct SEDs using contemporaneous X--ray and UV data observations. Instead, each UV/optical observation should be combined with the {\it average} X--ray spectrum over a time period just before the UV/optical observations, whose duration represents the disc response function, $\Psi(t)$,  at that particular wavelength.

The mass of the BH in NGC~6814 is $\sim10^7M_\odot$. According to Fig.\,15 in \cite{Kammoun_2021ApJ}, for such a BH with $a^\ast = 0$, accretion rate and 2--10 keV X--ray luminosity of 0.05 and 0.001 of the Eddington limit, respectively, the disc response function decreases by a factor of $\sim20$ (when compared with the maximum value) in the time period of $\sim1$ day at $\sim 2000$~\AA\ (second panel from left in their Fig. 15)  . As we find out from the spectral fits below, NGC~6814 may accrete at a rate higher than 0.05$L_{ \rm Edd}$ and the UVW1 band covers wavelengths longer than 2000~\AA.  
Consequently, the disc response function in NGC~6814 may be wider than $\sim1$~day, so, it would be better to use in each SED the average X--ray spectrum for a period longer than $\sim 1$~day prior to the UV observations\footnote{ We note that the width of the disc response function is always larger than the expected time--lag between X--rays and the disc variable emission in any spectral band, as the latter is equal to the centroid of the response function (see e.g. \S 4.1 in \citealt{Kammoun_2021ApJ} and references therein).}.
 
Therefore, we constructed one UV/X--ray SED from the XMM21 data, using only the last two data points in the UVW1 light curve (plotted in Fig. \ref{fig:lc}), and the average X--ray spectrum from the full EPIC--PN observation, which is $\sim 112$ ks (i.e. $\sim1.3$ days) long.
We utilized the task \textit{om2pha} for generating the UVW1 spectrum from the last two OM images. We also used one of the images corresponding to these last two data points to estimate the host galaxy contamination to the UV flux (see Appendix \ref{appendix2}) which we subtracted prior to the spectral fitting.
To extract the EPIC--PN spectrum, we again followed the standard SAS pipeline.

We also constructed 3 UV/X--ray SEDs from the \astrosat{} observation (AS1, AS2 and AS3, respectively) by considering only the 2 FUV data points just after 130~ks, 170~ks and 220~ks since the start of the \astrosat\ observation, and the average X--ray spectrum over the period between 0--130~ks, 40--170~ks and 90--220~ks 
(see Fig. \ref{fig:astrosat_lc}). The UV flux of the first and third \astrosat{} SEDs are comparable, while the UV flux is $\sim10\%$ higher in the second SED. The X--ray flux is highly variable in all periods before each UV flux, with no obvious pattern clearly associated with the UV points.
We used the {\sc uvittools} package to extract PHA files from FUV filter images, and we have utilized the \textsc{ftgrppha} command for optimal binning algorithm with a minimum of 25 counts per grouped bin \citep{Kaastra_2016}.

Before conducting spectral analysis, we applied aperture corrections to the UVIT/F154W band fluxes. We then subtracted the respective contributions from the host galaxy and emission lines from all four measured UV fluxes (see Appendix \ref{appendix1}, \ref{appendix2} and \ref{appendix3}).

%%%%%%%%%%%%%%%%%%%%%%%%%%%%%%%%%%%%%%%%%%%%%%%%%%%%%%%%%%%%%%%%%%%%%%%%%%%%%%%%%%%%
%-------------------- SECTION: TIMING ANALYSIS -----------------------------------%%
%%%%%%%%%%%%%%%%%%%%%%%%%%%%%%%%%%%%%%%%%%%%%%%%%%%%%%%%%%%%%%%%%%%%%%%%%%%%%%%%%%%%

\section{Timing Analysis} \label{sec:timing_analysis}

We calculated the fractional variability amplitude ($F_{\rm var}$)  of the light curves shown in Fig. \ref{fig:lc} using the formula as defined in \cite{Vaughan_2003MNRAS}.
The obtained $F_{var}$ results are listed in the Table \ref{tab:observation}.
The source is significantly variable in all the bands and the variability amplitude is higher in the X--ray band.

\begin{figure}[!ht]
\centering
\includegraphics[width=0.8\columnwidth]{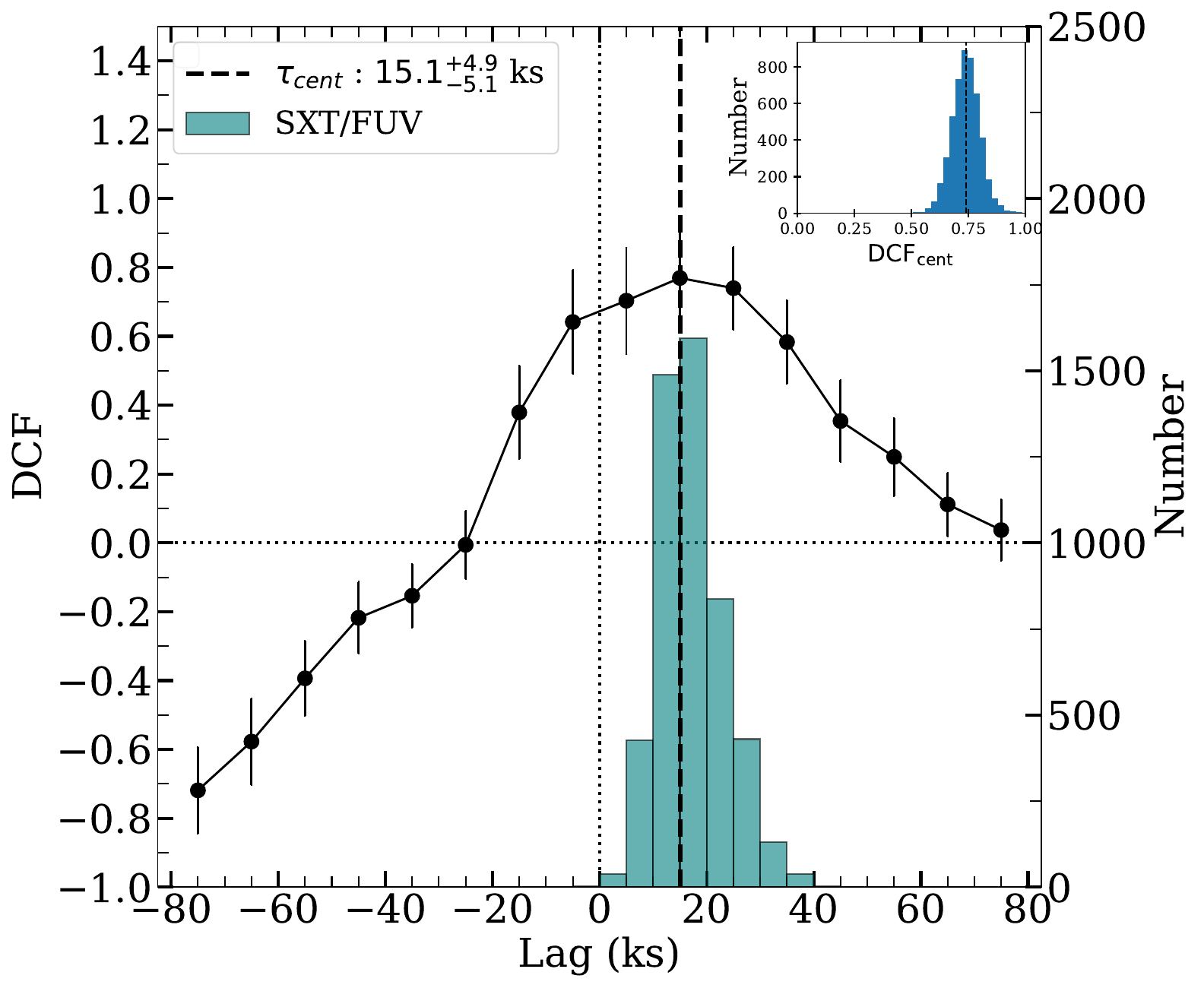}
\includegraphics[width=0.8\columnwidth]{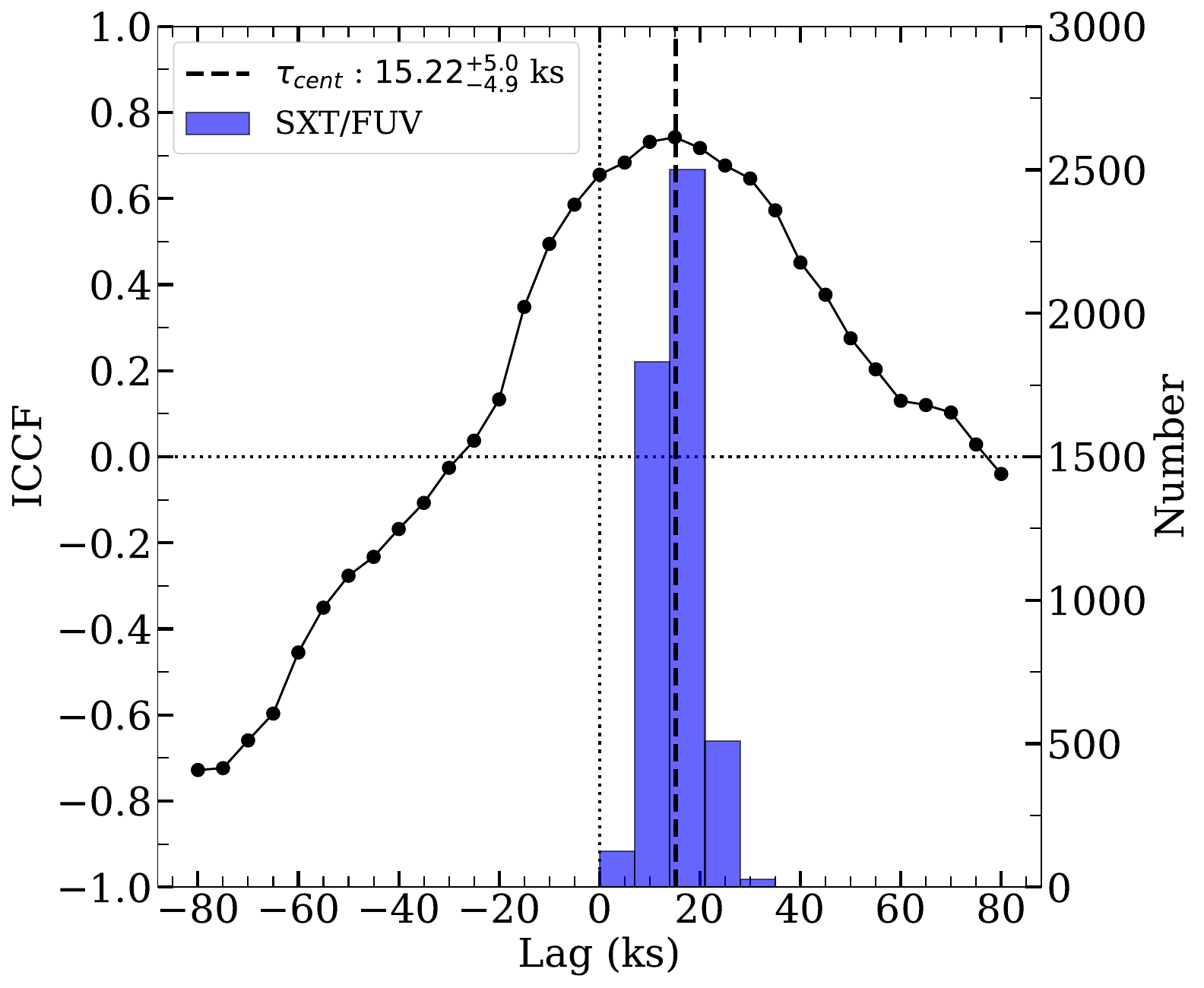}
\caption{\textit{Upper panel:} The SXT/FUV DCF curve (in black) for NGC~6814 using the \astrosat\, light curves. The histogram (in teal color) shows the centroid lag ($\tau_{\rm cent}$) distribution for 5,000 bootstrap realizations. The vertical dashed line indicates the mean of the  $\tau_{\rm cent}$ distribution. The inset (upper right corner) shows the $ DCF_{\rm cent}$ distribution for synthetic light curves. \textit{Lower panel:} Same for ICCF.}
\label{fig:astrosat_CCF_PLOTS}
\end{figure}

\begin{figure}[!ht]
\centering
\includegraphics[width=0.8\columnwidth]{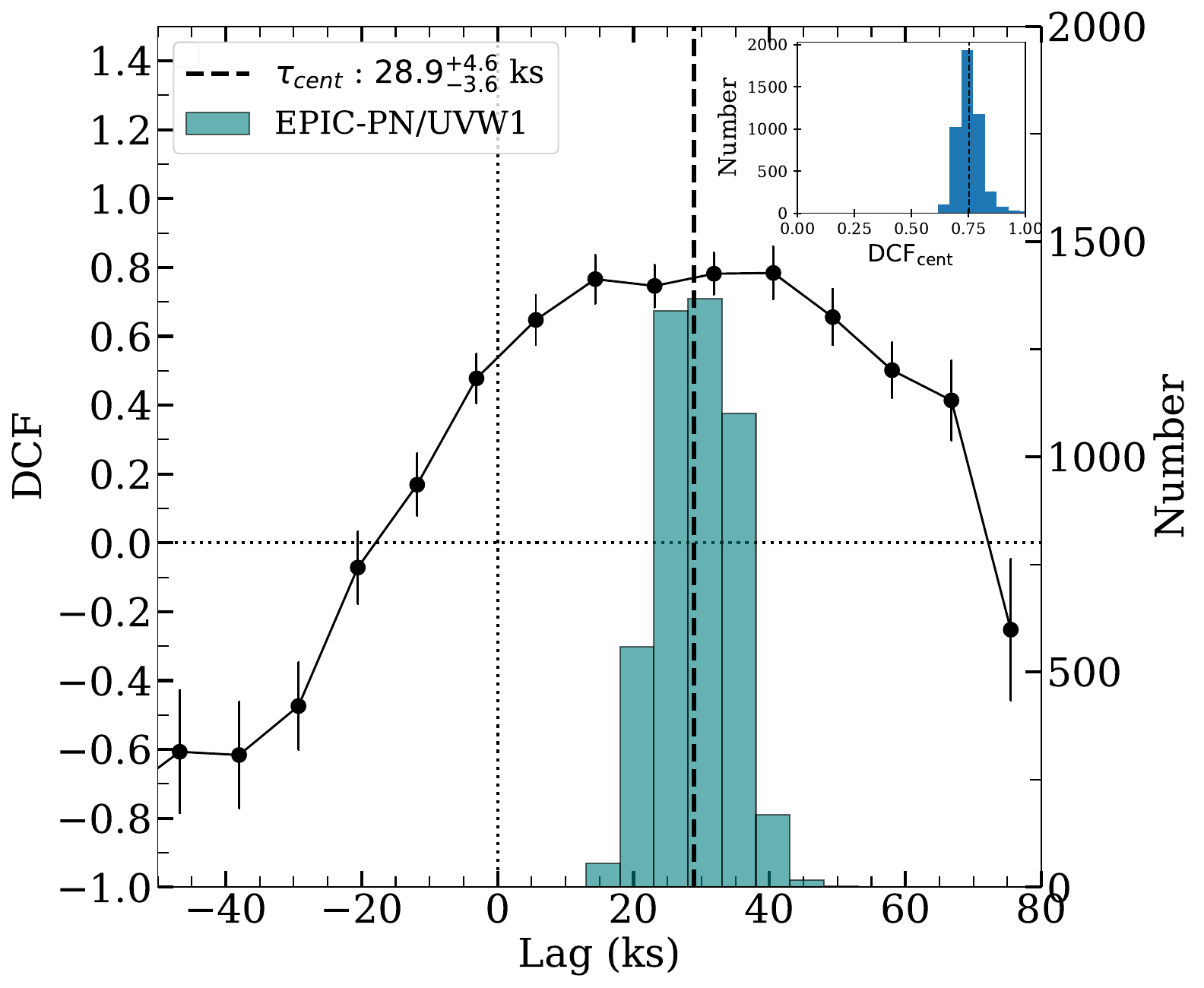}
\includegraphics[width=0.8\columnwidth]{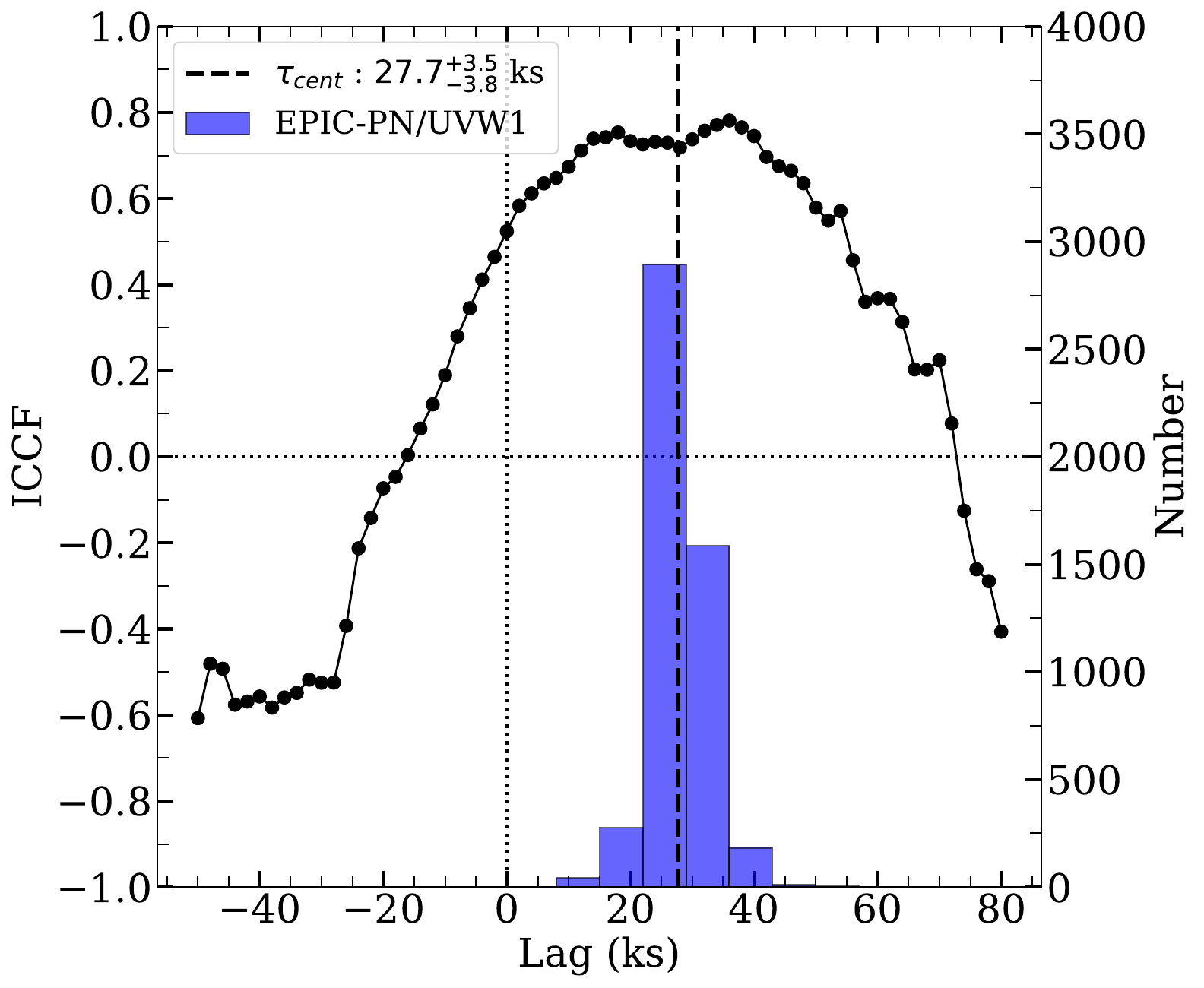}
\caption{Same as Fig. \ref{fig:astrosat_CCF_PLOTS} for XMM21 observation. }
\label{fig:xmm_CCF_PLOTS}
\end{figure}

We employed the Interpolated cross--correlation Function (ICCF; \citealt{peterson_1998PASP}) and discrete correlation Function (DCF; \citealt{edelson_1988ApJ}) techniques to compute the time--lags for the X--ray and UV light curves. For our analysis, we utilized Python-based packages: PyCCF \citep{Sun_2018} and PyDCF\footnote{\url{https://github.com/astronomerdamo/pydcf}}.
Further, to determine the lag distribution and associated uncertainties, we applied the random subset selection (RSS) technique to generate 5,000 pairs of bootstrap realizations. We present our cross--correlation results in Fig.~\ref{fig:astrosat_CCF_PLOTS} and \ref{fig:xmm_CCF_PLOTS}. In the upper panels, the DCF is shown as filled circles with error bars, while the centroid time--lag ($ \tau_{\rm cent}$) distribution is depicted in teal-colored histograms. In the lower panels, the ICCF results are illustrated with filled circles (without error bars) and blue histograms. The $\tau_{\rm cent}$ is determined by averaging the time--lags corresponding to CCF values $ >0.7~CCF_{\rm max}$. The histograms display the $ \tau_{\rm cent}$ distribution for all RSS-generated synthetic light curves, with the vertical dashed line indicating the mean of this distribution. Additionally, we compute the mean DCF at $ \tau_{\rm cent}$, denoted as $ DCF_{\rm cent}$, by averaging the DCF values  $ >0.7~DCF_{\rm max}$. The distribution of  $ DCF_{\rm cent}$ is shown in the inset of the upper panel of  Fig. \ref{fig:astrosat_CCF_PLOTS} and \ref{fig:xmm_CCF_PLOTS}. The mean of the  $ \tau_{\rm cent}$ distribution and the corresponding $ DCF_{\rm cent}$ values are reported in Table \ref{tab:timelag}.

%%%%%%%%%%%%%%%%%%%%%%%%%%%%%%%%%%%%%%%%%%%%%%%%%%%%%%%%%%%%%%%%%%%%%%%%%
%---------------Figure 5 : ASTROSAT CCF results -------------------
%%%%%%%%%%%%%%%%%%%%%%%%%%%%%%%%%%%%%%%%%%%%%%%%%%%%%%%%%%%%%%%%%%%%%%%%%

%%%%%%%%%%%%%%%%%%%%%%%%%%%%%%%%%%%%%%%%%%%%%%%%%%%%%%%%%%%%%%%%%%%%%%%%%
%---------------Figure 6: XMM-NEWTON CCF results -------------------
%%%%%%%%%%%%%%%%%%%%%%%%%%%%%%%%%%%%%%%%%%%%%%%%%%%%%%%%%%%%%%%%%%%%%%%%%

%%%%%%%%%%%%%%%%%%%%%%%%%%%%%%%%%%%%%%%%%%%%%%%%%%%%%%%%%%%%%%%%%%%%%%%%%%
%---------------------- Table 2: time--lag results ------------------------
%%%%%%%%%%%%%%%%%%%%%%%%%%%%%%%%%%%%%%%%%%%%%%%%%%%%%%%%%%%%%%%%%%%%%%%%%%

\begin{deluxetable}{cccc}
\tablecaption{Time-lags obtained using DCF and ICCF techniques. Errors correspond to 68\% confidence region for each parameter.}
\label{tab:timelag}
\tablehead{\colhead{} & \colhead{Mean $ \rm DCF_{\rm cent}$}  & \multicolumn{2}{c}{Mean $\rm \tau_{\rm cent} (ks)$}\\ 
\colhead{Filter} & \colhead{} & \colhead{DCF} & \colhead{ICCF}}
\colnumbers
\startdata 
  SXT/FUV & $0.74^{+0.06}_{-0.06}$ & $15.1^{+4.9}_{-5.1}$ & $15.2^{+5.0}_{-4.9}$ \\ \\
 EPIC--PN/UVW1 & $0.75^{+0.04}_{-0.05}$ &  $28.9^{+4.6}_{-3.6}$  & $27.7^{+3.5}_{-3.8}$  \\
\enddata
\end{deluxetable}

The mean $ DCF_{\rm cent}\sim 0.75$ indicates a good correlation between X--rays and the UV band variations in this source. The mean $ \tau_{\rm cent}$ for SXT/FUV and EPIC--PN/UVW1 is $\sim15$~ks ($\sim0.17$ days) and $\sim28$~ks ($\sim0.32$ days), respectively. The mean $ \tau_{\rm cent}$ in the UVW1 band is larger than the FUV band which indicates that the UVW1 photons ($\lambda_{\rm mean} = 2910$~\AA) originate from a larger disc area as compared to the FUV band ($\lambda_{\rm mean} = 1541$~\AA).
The time--lags calculated from both techniques are consistent within uncertainty and show that the UV photons are lagging the X--rays by their respective time--lag values.

%%%%%%%%%%%%%%%%%%%%%%%%%%%%%%%%%%%%%%%%%%%%%%%%%%%%%%%%%%%%%%%%%%%%%%%%%
%%%%%%--------------- Section: SPECTRAL Analysis-----------%%%%%%%%%%%%%
%%%%%%%%%%%%%%%%%%%%%%%%%%%%%%%%%%%%%%%%%%%%%%%%%%%%%%%%%%%%%%%%%%%%%%%%%

\section{Spectral Analysis} \label{sec:spec_analysis}

\subsection{Fitting Procedure}

We fitted the 4 X--ray and UV spectra using {\sc xspec} (version: 12.12.1; \citealt{Arnaud_1996ASPC_xspec}). The main model component is \textsc{kynsed}, but we also need to consider additional model components in order to fit the energy spectra well. We describe below these components, before reporting the best--fit results.

{\it Galactic absorption:} 
We used the \textsc{tbabs} model to fit the Galactic absorption in the X--ray band. We fixed the Galactic column density at $N_H = 8.42 \times 10^{20} cm^{-2}$  \citep{HI4PI_2016}. We used the \textsc{redden} model component to take into account the Galactic dust extinction in the UV/optical band using the extinction law by \cite{Cardelli_1989ApJ_redden}. We calculated the V-band extinction ($ A_V$) using $N_H$, and the gas-to-dust ratio $N_H = 2.21\times 10^{21} \times A_V$ \citep{Guver_2009MNRAS_extinction}. We obtained $A_V = 0.38$, which matches the previous values used by \cite{winkler_1997MNRAS_extinction} and \cite{Allan_2016MNRAS_extinction} within uncertainty. Then we used the ratio of total to selective extinction, $R_V = 3.1$ to get the extinction $E(B-V) = 0.123$. We fixed the galactic reddening parameter for the UV spectra at this value in all the model fits we performed.

{\it Absorption in the host galaxy:} We found strong evidence for absorption in the soft X--ray band, in addition to the one due to the Galactic absorption. We tested for the presence of neutral, ionized and warm absorbers using the models \textsc{ztbabs/tbpcf}, \textsc{zxipcf}, and  \textsc{xabs} \citep{steenburgge_2003_xabs} models, respectively.  The best--fit $\chi^2$ values were systematically smaller in the case when we assumed absorption by a partially covering neutral absorber. The best fits with the use of \textsc{xabs} were almost similarly good; however, the best fit ionization parameter was always very small ($log~\xi < 0$) indicating the presence of a neutral absorber instead of an ionized one. \cite{Kang_2023MNRAS} also pointed out the absence of a warm absorber in the XMM21 observation. Thus we adopted the spectral component \textsc{tbpcf} to account for the absorption in the host galaxy, in the X--ray band. We also added a \textsc{zredden} component to account for the associated extinction in the UV band. We kept the respective reddening parameters, $E(B-V)_{\rm host}$, free to vary during the model fits. 

{\it X--ray reflection from neutral material:} During various model fits the data, we also noticed the presence of narrow Fe~K$\alpha$ (6.4 keV) and K$\beta$ (6.94 keV) lines in the XMM21 observation. Given the energy and width of the observed lines, we assumed that it arises from X--ray reflection from neutral material, so we considered the neutral reflection model \textsc{pexmon} \citep{Nandra_2007MNRAS_PEXMON}, for the model fitting to the XMM21 data only. The inclination angle in \textsc{pexmon} was not constrained, so we fixed it at $\theta_p = 45^\circ $ in all cases.

The final model for the simultaneous fitting of all the UV/X--ray spectra is: 

%\begin{multline}
%{\rm Model} =  \textsc{redden} \times 
%    \textsc{zredden} \times \textsc{tbabs} \times \textsc{tbpcf} \times  \\
%    (\textsc{kynsed} + \textsc{pexmon})
%\label{eq:model}.
%\end{multline}

\begin{eqnarray}
    {\rm Model} =  \textsc{redden} \times 
    \textsc{zredden} \times \textsc{tbabs} \times \textsc{tbpcf} \times \nonumber \\
    (\textsc{kynsed} + \textsc{pexmon})
\label{eq:model}.
\end{eqnarray}

\textsc{kynsed} is a physical model for the broadband SED of X--ray illuminated accretion discs that take into account all the relativistic effects on light propagation and on the transformation of the photon energy as well as the mutual interaction of the accretion disc and the X--ray corona \citep{Dovciak_kynsed2022}.  The model assumes a Keplerian disc co-rotating the central BH and an X--ray source in the lamp--post geometry. The X--ray corona emits isotropically in its rest--frame. The model includes disc heating due to the thermalisation of the absorbed part of the X--rays which illuminate the disc and also includes X--ray reflection due to the disc illumination.

BH mass and accretion rate are the main physical parameters of the model. We fixed the BH mass at $M_{\rm BH}=1.09\times 10^7 M_\odot$, and we kept the accretion rate the same while fitting all four spectra. This implies that we wish to investigate whether the observed UV variations are entirely due to the X--ray variations. We fixed the high--energy cut--off to be 200 keV \citep{Malizia_2014ApJ} and the outer disc radius at $R_{\rm out} = 10^4~r_g$. The model normalization parameter is defined as $ 1/D_L^2$, where $ D_L$ is the luminosity distance in Mpc. We have assumed that  $ D_L = 21.65$~ Mpc \citep{Bentz_2019ApJ_cepheid_distance} and kept this parameter fixed while fitting the four SEDs. 

The remaining physical parameters of the model are: the BH spin, $a^{\ast}$, the color correction factor, $f_{\rm col}$\footnote{This is a constant that is frequently used to take into account the spectral hardening of the disc emission due to photon scattering off free electrons in the upper disc layers.}, the corona height, $h_c$, the photon index $\Gamma$, the disc inclination, $\theta$, and the power given to the corona, $L_{\rm transf}$. If the $L_{\rm transf}$ is positive, it is equal to the power that is released by the accretion process below a `transition' radius, $R_t$, and then it is transferred to the
corona by an unknown physical mechanism. If this parameter is negative, then it is equal to the power that is given to the corona from an unknown external source.

%%%%%%%%%%%%%%%%%%%%%%%%%%%%%%%%%%%%%%%%%%%%%%%%%%%%%%%%%%%%%
%%%%%%%--------------------- TABLE 3 ------------------------
%%%%%%%%%%%%%%%%%%%%%%%%%%%%%%%%%%%%%%%%%%%%%%%%%%%%%%%%%%%%%

\begin{deluxetable}{ccccc}
\tablecaption{Best--fit $\chi^2_{\rm min}$ values for 418 degrees of freedom and `evidence ratio' ($\epsilon$) when fitting the four SEDs having fixed the inclination angle to $70^\circ$ in the \textsc{kynsed} model, for different BH spins (from 0.0 to 0.7), negative/positive $L_{\rm transf}$ and $f_{\rm col}$ (-1 means the default prescription used in \citealt{Done_2012MNRAS}). %The BH spin of 0.998 was giving non-converging fits. 
The best--fit $\chi^2$ values for pairs of (BH spin, $f_{\rm col}$) values which fitted the data as good as the one which provided the smallest $\chi^2$ (i.e. pairs with $\epsilon> 0.1$) are indicated in bold font.}
\label{tab:fit_chisquare}
\tablehead{
 \colhead{$f_{\rm col}$} & \multicolumn{4}{c}{BH Spin, $a^{\ast}$} \\
 \colhead{} & \colhead{0.0} & \colhead{0.3} & \colhead{0.5} & \colhead{0.7}
}
\colnumbers
\startdata 
\multicolumn{5}{c}{$\rm L_{\rm transf}<0$} \\\\
 -1 & 467.0 & 471.1  &  475.4 &  481.7 \\
 %----------------------------------------------------
&  $\mathbf{\epsilon=0.18}$ & $\epsilon=0.02$ & $\epsilon=0$ & $\epsilon=0$\\
 1  & 467.9 &  469.6 & 469.5 &  470.0 \\
 %----------------------------------------------------
 &  $\mathbf{\epsilon=0.12}$ & $\epsilon=0.05$ & $\epsilon=0.05$ & $\epsilon=0.04$\\
 1.7 &  463.6 & 465.3 &  466.3 & 469.3 \\
 %----------------------------------------------------
 &  $\mathbf{\epsilon=1}$ & $\mathbf{\epsilon=0.43}$ & $\mathbf{\epsilon=0.26}$ & $\epsilon=0.06$\\
 %----------------------------------------------------
  2.4 & 463.8 & 466.8 & 470.33 &  475.9 \\
  %----------------------------------------------------
  &  $\mathbf{\epsilon=0.90}$ & $\mathbf{\epsilon=0.20}$ & $\epsilon=0.03$ & $\epsilon=0$\\
 \hline
 %----------------------------------------------------
 \multicolumn{5}{c}{ $\rm L_{\rm transf}>0$}\\\\
  -1 & 467.4 & 470.8 & 472.7  & 479.8\\
  %----------------------------------------------------
 &  $\mathbf{\epsilon=0.35}$ & $\epsilon=0.06$ & $\epsilon=0.02$ & $\epsilon=0$\\
 %----------------------------------------------------
 1 & 471.0 & 472.5 &  475.3  & 479.2 \\
 %----------------------------------------------------
 &  $\epsilon=0.06$ & $\epsilon=0.03$ & $\epsilon=0$ & $\epsilon=0$\\
 %----------------------------------------------------
  1.7  & 465.3 &  466.9 &  468.9   & 471.7  \\
  %----------------------------------------------------
  &  $\mathbf{\epsilon=1}$ & $\mathbf{\epsilon=0.45}$ & $\mathbf{\epsilon=0.17}$ & $\epsilon=0.04$\\
  %----------------------------------------------------
  2.4  &  467.6 & 468.4 &  469.0  & 473.9  \\ 
  %----------------------------------------------------
  &  $\mathbf{\epsilon=0.32}$ & $\mathbf{\epsilon=0.21}$ & $\mathbf{\epsilon=0.16}$ & $\epsilon=0.01$\\
\enddata
\end{deluxetable}

%\underline{\it Initial model fits.}
The number of physical parameters in \textsc{kynsed} is large and they cannot all be constrained by fitting the four SEDs. We therefore decided to initially fit the data using a grid of spin, inclination, and $f_{\rm col}$ values. We considered the values of $a^{\ast} =$ 0, 0.3, 0.5, 0.7 and 0.998, $\theta = 20^\circ $, $40^\circ$, $60^\circ$ and $70^\circ$, and $f_{\rm col} \rm =-1$\footnote{$f_{\rm col} = -1$ is the default value of \textsc{kynsed}, and uses the prescription of \cite{Done_2012MNRAS} for the correction of the emitted disc spectrum due to photon scattering effects on the disc atmosphere.}, 1, 1.7, and 2.4. 
We fitted the data by linking $L_{\rm transf}$ for all three SXT spectra (since they were taken with a short time difference). We also linked all the relevant \textsc{kynsed} parameters in the UV spectrum with the corresponding parameters for the X--ray spectrum, and we added a 2\% systematic to the model. 

The model fits the data best for $\theta=70^\circ$. 
We also examined the effect of fixing the inclination angle to different values. When the inclination is fixed to $69^\circ$, $68^\circ$, $67^\circ$, $66^\circ$, $65^\circ$, and $60^\circ$, the $\chi^2$ values systematically worsen to 466.9, 467.9, 469.2, 470.5, 472.0, and 479.6, respectively.
For a single parameter of interest, a change in $\chi^2$ of 6.63 corresponds to a 99\% confidence limit. Based on this, inclination angles in the range of approximately $ 66^\circ-70^\circ$ are statistically acceptable within the 99\% confidence level.
While fixing the inclination to a value slightly higher than $70^\circ$ improves the fit, such high inclinations would not be physically meaningful for a source classified as a type 1.5 Seyfert, which is expected to be observed at intermediate viewing angles. 
Our adopted inclination angle is consistent with the disc inclination of $\sim 67^{\circ}$ obtained by \cite{Gallo_ngc6814_2021ApJ}.
\cite{Rosenblatt_1994ApJ} also predicted a high disc inclination for this source, between $\sim 60^\circ$--$80^\circ$ based on the H$\beta$ line profile variations. Furthermore, \cite{Pancoast_2014MNRAS} found that the BLR geometry in this source may be that of a wide thick disc with an inclination of $49.4^{+20.4\circ}_{-22.2}$, which matches with our accretion disc inclination (within the errors which, arguably, are large), assuming that the BLR disc is located above the accretion disc in NGC~6814. Additionally, the clumpy torus modelling of NGC~6814, as presented in \cite{Ramos_2011ApJ_torus}, suggests a relatively narrower half angular width of the torus ($\sigma \approx 25^\circ$) and a moderate inclination angle which is consistent with our adopted value of inclination angle within uncertainty.

Table \ref{tab:fit_chisquare} lists the best--fit model results (i.e. the $\chi^2_{\rm min}$ values) in the case when we keep $\theta$ fixed at $70^\circ$, for the various BH spins, $L_{\rm trasnf}$, and $f_{\rm col}$ values. We do not list $\chi^2_{\rm min}$ values for $a^{\ast}=0.998$, because the fit would not converge in most of the cases when assuming this spin, or $\chi^2_{\rm min}$ would be very large. In an attempt to decide which one of the ($a^{\ast},~f_{\rm col},~ L_{\rm transf}$) combination fits best the data, we calculate the Akaike information criterion (AIC) and the `evidence ratio' ($\epsilon$) for each model.  
The `evidence ratio' is defined as $\epsilon = e^{- \frac{\Delta[\rm AIC_{min,i}]}{2}}$ (\citealt{Akaike1973InformationTA}, \citealt{Emmanoulopoulos_2016MNRAS}), where, for our case, $\Delta[\rm AIC_{min,i}]$ is the difference between $\chi^2_{\rm min}$ of the models that resulted in the smallest $\chi^2_{\rm min}$ value, and the $i^{th}$ model. The evidence ratio measures the relative likelihood of the model with the smallest $\chi^2_{\rm min}$ versus all the other models. For example, a value of $\epsilon=0.1$ indicates that the $i^{th}$ model is 10 times less probable than the model which provided the smallest $\chi^2_{\rm min}$ to be the correct model. 

%%%%%%%%%%%%%%%%%%%%%%%%%%%%%%%%%%%%%%%%%%%%%%%%%%%%%%%%%%%%%%%%%%%%%%%
%------------- Figure 7: best--fit SED
%%%%%%%%%%%%%%%%%%%%%%%%%%%%%%%%%%%%%%%%%%%%%%%%%%%%%%%%%%%%%%%%%%%%%%%

\begin{figure*}[!ht]
\centering
\includegraphics[width=1.8\columnwidth]{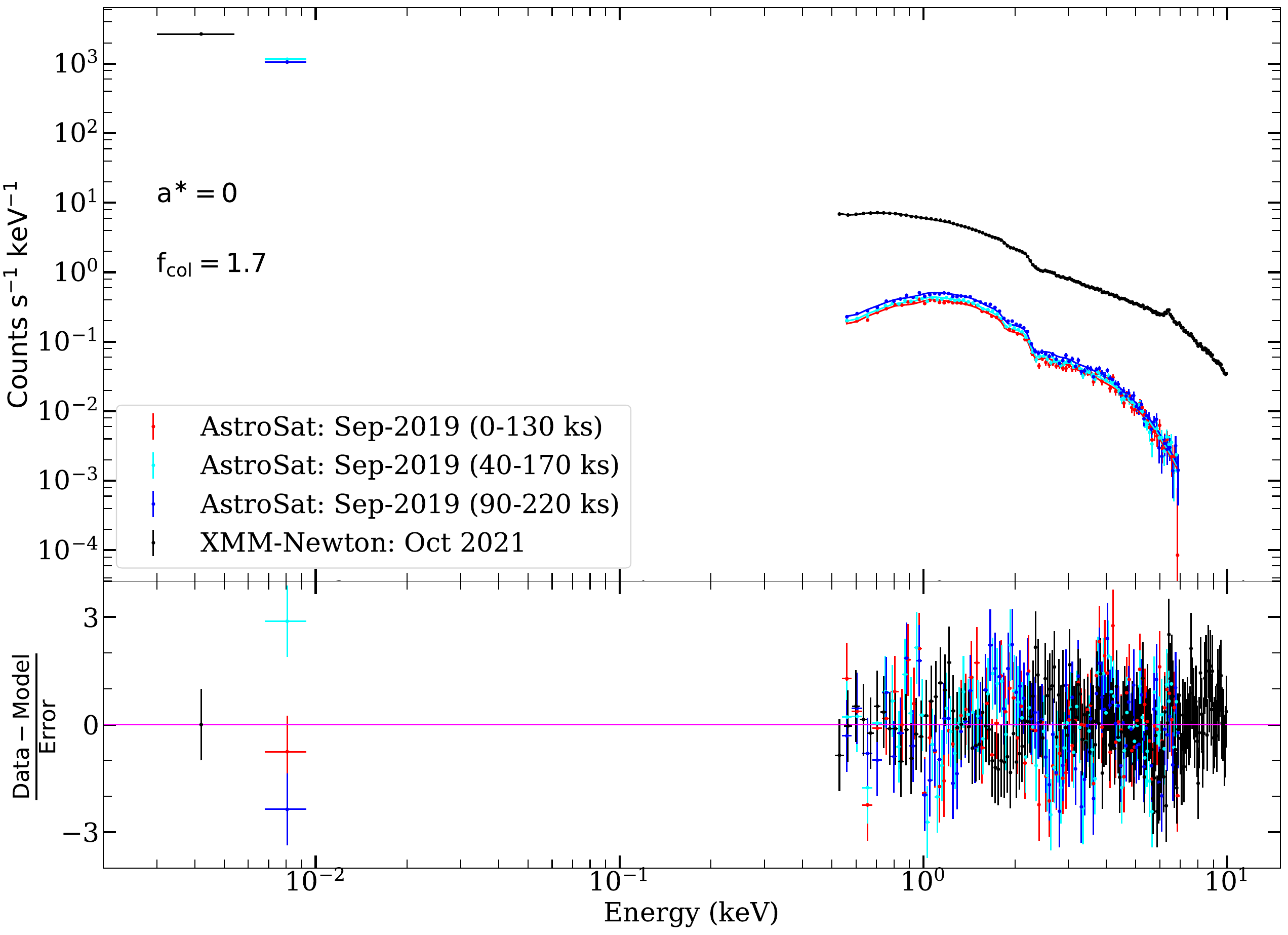}
\caption{\textit{Upper panel:} Time-resolved SEDs of NGC~6814 fitted with the best--fit model: $ \textsc{redden} \times \textsc{zredden} \times \textsc{tbabs}   \times \textsc{tbpcf} \times (\textsc{kynsed} + \textsc{pexmon})$ for the case $\rm a^{\ast}=0$, $\rm f_{\rm col}=1.7$ and $\rm L_{\rm transf}>0$. Table \ref{tab:fit_val} lists the best--fit model parameters. \textit{Lower panel:} the delchi plot for the best--fit model.}
\label{fig:SEDs}
\end{figure*}

%%%%%%%%%%%%%%%%%%%%%%%%%%%%%%%%%%%%%%%%%%%%%%%%%%%%%%%%%%%%%%%%%%%%%%%
%------------- Figure 8: best--fit SED
%%%%%%%%%%%%%%%%%%%%%%%%%%%%%%%%%%%%%%%%%%%%%%%%%%%%%%%%%%%%%%%%%%%%%%%

\begin{figure}
    \includegraphics[width=1\columnwidth]{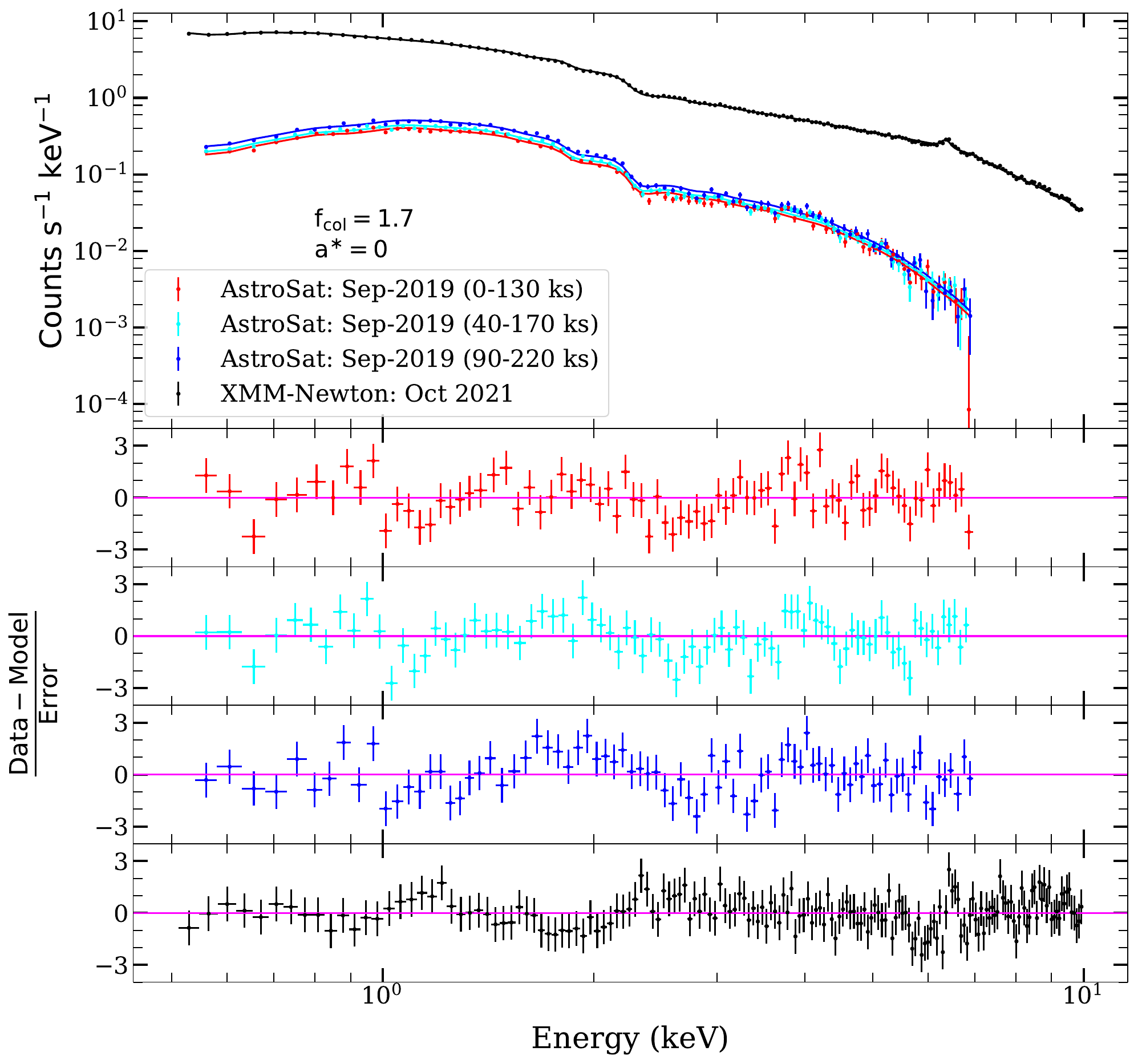}
\caption{Same as in Fig.\,\ref{fig:SEDs}, but only for the X--ray spectra to show clearly the agreement between the best model fit and the X--ray SEDs.}   
\label{fig:Xrayspectra}
\end{figure}

The model combination which provides the smallest $\chi^2_{\rm min}$ is the one with $a^{\ast}=0, f_{\rm col}=1.7$, and $L_{\rm transf}>0$. An almost equally good fit is provided by the same spin, in the case of an externally powered corona ($L_{\rm transf}<0$), and $f_{\rm col}= 1.7, ~2.4$. Good fits are also provided by models with $f_{\rm col}=1.7$ and $a^{\ast}=0.0,~ 0.3$ and 0.5, for an externally powered corona and for a corona powered by the accretion process, since in all these cases $\epsilon>0.1$. In addition, we can also get good fits assuming $f_{\rm col}=-1$, for spin 0. 

Our results suggest that we cannot constrain $f_{\rm col}$ by fitting the four SEDs we study in this work. On the other hand, we find that the BH spin is $\leq 0.5$ in all cases with $\epsilon>0.1$. In other words, if we allow BH spin to be larger than 0.5, then the resulting model fits are more than 10 times less probable, when compared with the best--fit model. We cannot be certain whether the BH is nonrotating or not, but our results indicate that it is probably not rapidly spinning. 

%%%%%%%%%%%%%%%%%%%%%%%%%%%%%%%%%%%%%%%%%%%%%%%%%%%%%%%%%%%%%
%%%%%%%--------------------- TABLE 4 ------------------------
%%%%%%%%%%%%%%%%%%%%%%%%%%%%%%%%%%%%%%%%%%%%%%%%%%%%%%%%%%%%%

\begin{deluxetable*}{ccccccc}
\tablecaption{Best--fit results ($\chi^2 = 465.3/418$) for  $L_{\rm transf}>0$,  inclination angle fixed to $70^{\circ}$, BH spin $a^{\ast}=0$ and $f_{\rm col} = 1.7$ in the \textsc{kynsed} model.
Errors representing a 90\% confidence limit for a single parameter. Values in brackets represent those obtained without accounting for the average emission line contribution in the UV flux measurements.}
\label{tab:fit_val}
\tablehead{
\colhead{Model} & \colhead{Parameters} & \colhead{Units} &  \colhead{ AS1} & \colhead{AS2} & \colhead{ AS3} & \colhead{XMM21} \\
}
\colnumbers
\startdata 
 KYNSED & Spin, $\rm a^{\ast}$ &  & \multicolumn{4}{c}{0 (fixed)}\\\\
 %----------------------------------------------------
 & arate & ($\rm \times 10^{-2} L_{\rm Edd}$) & \multicolumn{4}{c}{$10.53^{+0.15}_{-0.10}$}\\
 & & &  \multicolumn{4}{c}{(9.34)} \\\\
 %----------------------------------------------------
 & $\rm L_{\rm transf}$ & ($\rm L_{\rm transf}/L_{\rm disc}$) & \multicolumn{3}{c}{$0.19_{-0.02}^{+0.02}$} & $0.09_{-0.00}^{+0.01}$\\
 & &  & \multicolumn{3}{c}{(0.21)} & (0.11) \\\\
 %-----------------------------------------------------
 & height, $\rm h_c$ & ($\rm r_g$) & $9.09^{+2.53}_{-1.74}$ & $12.58^{+5.63}_{-3.14}$ & $23.78^{+12.33}_{-7.53}$ & $\geq 27.97 $\\
 & & & (9.27)  & (13.13)  & (24.46) &  ( $\geq 27.62$) \\\\
  %-----------------------------------------------------
 & Photon Index, $\Gamma$ & & $1.80^{+0.02}_{-0.02}$ & $1.81^{+0.02}_{-0.02}$ & $1.86^{+0.02}_{-0.03}$ & $1.88^{+0.02}_{-0.03}$\\
 & & &   (1.80) & (1.81)  &  (1.86)  &  (1.88)  \\\\
  %-----------------------------------------------------
 PEXMON & norm & ($\times 10^{-3}$) & \multicolumn{3}{c}{-} & $2.48^{+0.51}_{-0.72}$\\
 &&&  \multicolumn{3}{c}{-} &  (2.46) \\\\
  %-----------------------------------------------------
 TBPCF &  $\rm N_H$ & ($\rm 10^{22}~cm^{-2}$) & \multicolumn{3}{c}{$23.41^{+7.98}_{-5.14}$} & $1.05^{+0.25}_{-0.24}$\\
 &&&  \multicolumn{3}{c}{(22.04)} &  (1.05)  \\\\
  %-----------------------------------------------------
 &  $\rm f_{\rm cov}$ &  & \multicolumn{3}{c}{$0.24^{+0.05}_{-0.06}$} & $0.19^{+0.03}_{-0.04}$\\
 &&&  \multicolumn{3}{c}{(0.22)} &  (0.19) \\\\
  %-----------------------------------------------------
 ZREDDEN &$\rm E(B-V)_{\rm host}$ & & \multicolumn{3}{c}{$0.24^{+0.01}_{-0.01}$} & $0.35^{+0.01}_{-0.01}$\\
 & & &   \multicolumn{3}{c}{(0.21)} &  (0.29) \\\\ \hline
  %-----------------------------------------------------
 & $\rm F_{2-10keV}$ & ($\rm \times 10^{-11}ergs~s^{-1}cm^{-2}$) & 4.4 & 4.7  & 5.2  & 2.8 \\
 & $\rm L_{2-10keV}$  & ($\rm \times  10^{-3} L_{\rm Edd}$) &  1.8 & 1.9  & 2.1 & 1.1 \\
 & $\rm{R_c}$   & ($\rm r_g$) & 5.8 & 6.3 & 8.6 & 9.3\\
 & $\rm R_{t}$  & ($\rm r_g$) & \multicolumn{3}{c}{14.2} & 11.1 \\ 
\enddata
\end{deluxetable*}

\subsection{Best--Fit Results} \label{sec:bestfit_results}

We fitted the 4 UV/X--ray SEDs simultaneously with the model defined by Eq.~\ref{eq:model} in the case where $a^{\ast}=0$, $f_{\rm col}=1.7$ and $L_{\rm transf}>0$. Although we cannot constrain $f_{\rm col}$, and all we can say about the BH spin is that it is smaller than 0.5, we fixed the BH spin at zero and $f_{\rm col}$ at 1.7 because this combination gave the smallest $\chi^2_{\rm min}$ value among all the models we considered in the previous section (see Table \ref{tab:fit_chisquare}).

The model fits well the data with $\chi^2_{\rm min}=465.3/418$ degrees of freedom (dof; $p_{\rm null}=0.06$). Table \ref{tab:fit_val} lists the best--fit results, and we also list in brackets the best--fit values obtained without accounting for the average emission line contributions in the respective UV filter bands (see Appendix \ref{appendix3} for calculation).
Fig.\,\ref{fig:SEDs} shows the best--fit model, together with the best--fit residuals. Since the energy range in this figure is large, it is not easy to show the quality of the model fit to the X--ray data. For that reason, Fig. \ref{fig:Xrayspectra} shows the best--fit model to the X--ray spectra only, in order to demonstrate clearer the quality of the model fit to the X--ray spectra.

We tested the effect of different $R_{\rm out}$ on our best--fit results. We found that the best--fit parameters and $\chi^2$ values do not change significantly even for an outer disc radius as small as $1000~r_g$. 
\textsc{kynsed} calculates what should be the radius of the corona ($R_c$) to explain the observed X--ray luminosity, using the conservation of energy and photon numbers during the Comptonisation process. This estimation can be used to check the model's validity since the height of the corona should always be larger than the sum of its radius and the radius of the event horizon. We obtained the corona radius using the \textsc{xset} command, and the relevant values are listed in the bottom rows of Table \ref{tab:fit_val}. These estimates are acceptable, as the best--fit corona heights are always larger than $R_c + 2r_g$ (where $2r_g$ is the horizon radius for $a^{\ast}=0$).

Our best--fit results indicate that the accretion power transferred to the corona is $\sim 20\%$ and $10\%$ of the total power that is released by the accretion process in the disc for the \astrosat{} 2019 and the XMM21 observations, respectively. The corresponding transition radius are $\sim14$ and $\sim 11~r_g$ (see Table \ref{tab:fit_val}). The best--fit model parameters and quality are similar even in the case when $L_{\rm transf}$ is negative.

Our results indicate that it is possible that the accretion rate in this source remained constant between (and obviously during) the \astrosat{} and XMM21 observations and that the observed UV variations in this source could be entirely due to X--ray reverberation. We find that $\dot{m}/\dot{m}_{\rm Edd}\sim 0.1$, which gives a bolometric luminosity of $L_{\rm bol}\sim 1.4\times 10^{44}~ergs~s^{-1}$. This is $\sim3$ times larger than the value quoted in \cite{Duras_2020_bolometric}. 

We see an increasing trend in the height of the corona for \astrosat{}~ 2019 observation ranging from $\sim 10$ to $25~ r_g$ (column 4-6 in Table \ref{tab:fit_val}). For the XMM21 spectrum, we only get a lower limit on corona height, i.e. $h_c \geq 28~r_g$. We cannot investigate whether there is a correlation between $h_c$ and $L_{\rm transf}$ since we kept this value the same for the three \astrosat{} SEDs. However, there seems to be a positive correlation between $\Gamma$ and $h_c$, although not a strong one, given the errors on $\Gamma$. Even if real, this correlation should be weak.

We note the presence of a neutral absorber at the host galaxy with column density $N_H\sim 10^{22}~cm^{-2}$ and covering fraction $f_{\rm cov}\sim0.2$ for XMM21 observation. The obtained $N_H$ is an order of magnitude higher ($\sim 10^{23}~cm^{-2}$) for \astrosat{} data. This may suggest a possible change in the column density of the neutral absorber over the years. Previous studies have proposed that such variable absorption may originate from clumpy material near the broad-line region (BLR) or torus rather than from large-scale galactic absorption (see for e.g. \citealt{Ramos_2017NatAs, Shrabani_2024ApJ_NGC4151}). The substantial difference in $N_H$ between the observations further supports this scenario. We found an average host galaxy reddening $E(B-V)_{\rm host} \sim 0.3$, which is consistent with the previously adopted value for the nucleus reddening in NGC~6814 by \cite{winkler_1997MNRAS_extinction}.

%Furthermore, the detection of the \textsc{pexmon} component in the \xmm{} data but not in the \astrosat{} spectrum suggests two possibilities: either the distant reflection is intrinsically weak or highly variable. Alternatively, its absence in the \astrosat{} data may be due to the instrument’s lower effective area and signal--to--noise ratio, making the feature undetectable rather than truly absent.

Furthermore, the detection of the \textsc{pexmon} component in the \xmm{} data but not in the \astrosat{} spectra suggests that the distant reflector may be variable. However, it is also possible that we do not detect this component in the \astrosat{} data because of the instrument's lower effective area compared to \xmm{}. In fact, when we add a \textsc{pexmon} component in the \astrosat{} model with its parameters fixed to the best--fit values from the \xmm{} analysis, the fit improves slightly ($\Delta\chi^2=0.7$), although the improvement is not highly significant. This result indicates that the same \textsc{pexmon} component can be present during the \astrosat{} observations, but we cannot constraint it with the \astrosat{} SEDs.

%%%%%%%%%%%%%%%% SECTION: SPECTRAL-TIMING CONNECTION---------------

\section{Spectral--Timing connection}\label{sec:spectral_timing}

Having fitted the time-resolved SEDs with \textsc{kynsed}, we can now test whether X--ray reverberation can also fit the observed time--lags of the source with the same physical parameters that provided a good fit to the energy spectra. We can do this by using the analytic expression of \cite{Kammoun_2021ApJ}, taking into account the updated analytic expression (see Eq. B1 and B2 in \citealt{kammoun_lag_2023MNRAS}) for computing time--lags which also incorporates the dependence on $f_{\rm col}$. 

The \cite{Kammoun_2021ApJ} models were developed by assuming that X--rays are powered by a source which is not associated with the accretion process.  We fitted the energy spectra by assuming that the X--ray source is powered by the accretion process itself. However, \cite{kammoun_lag_2023MNRAS} demonstrated (see Fig. B3 in their paper) that, irrespective of whether the corona is powered externally or by the accretion process, the time--lag difference can be smaller than 20\% for $L_{\rm transf}/L_{\rm disc}<~0.4$, as is the case with the NGC~6814 (according to our best--fit results).

\begin{figure}[!ht]
\centering
\includegraphics[width=1\columnwidth]{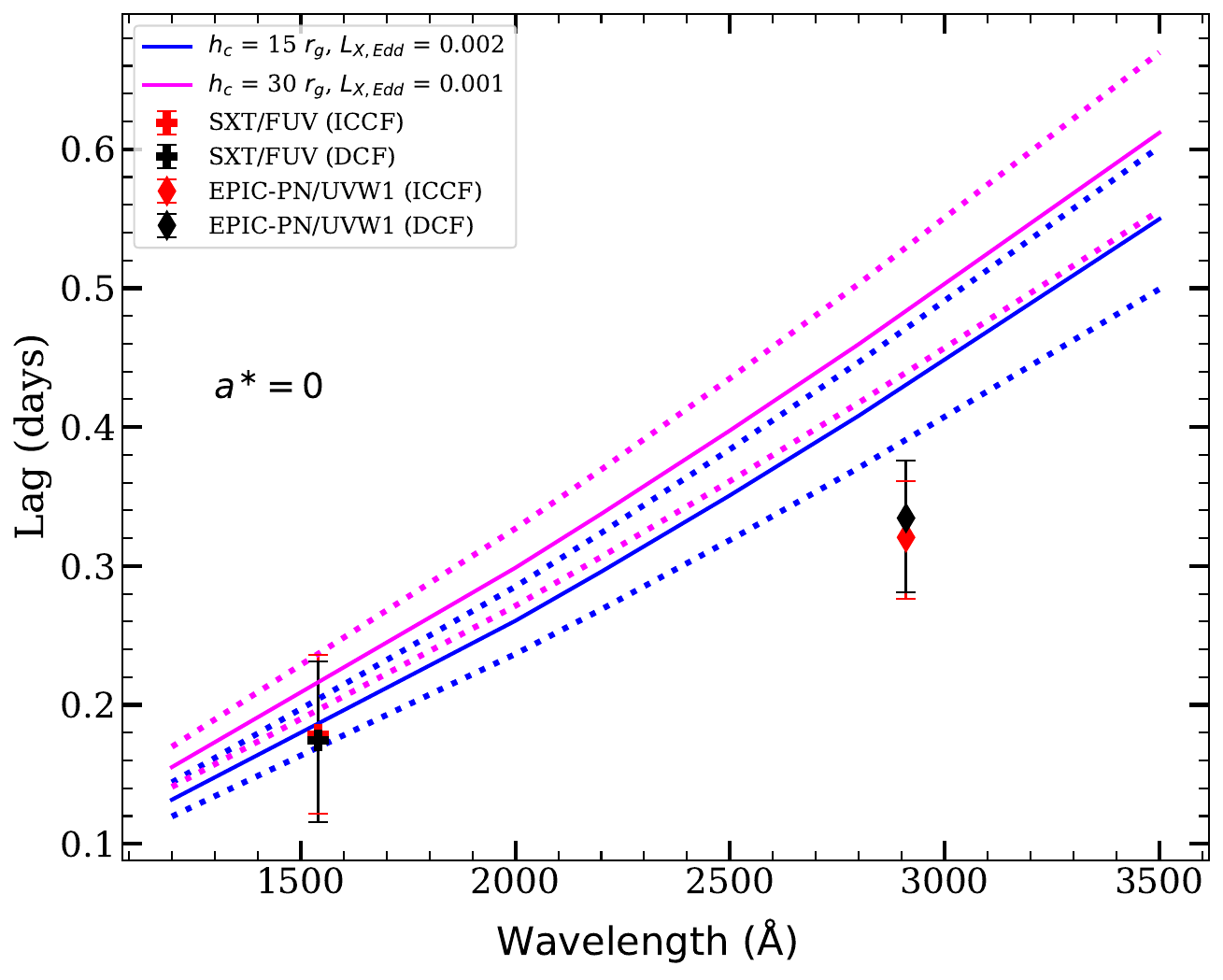}%\label{fig:my_lag}

\includegraphics[width=1\columnwidth]{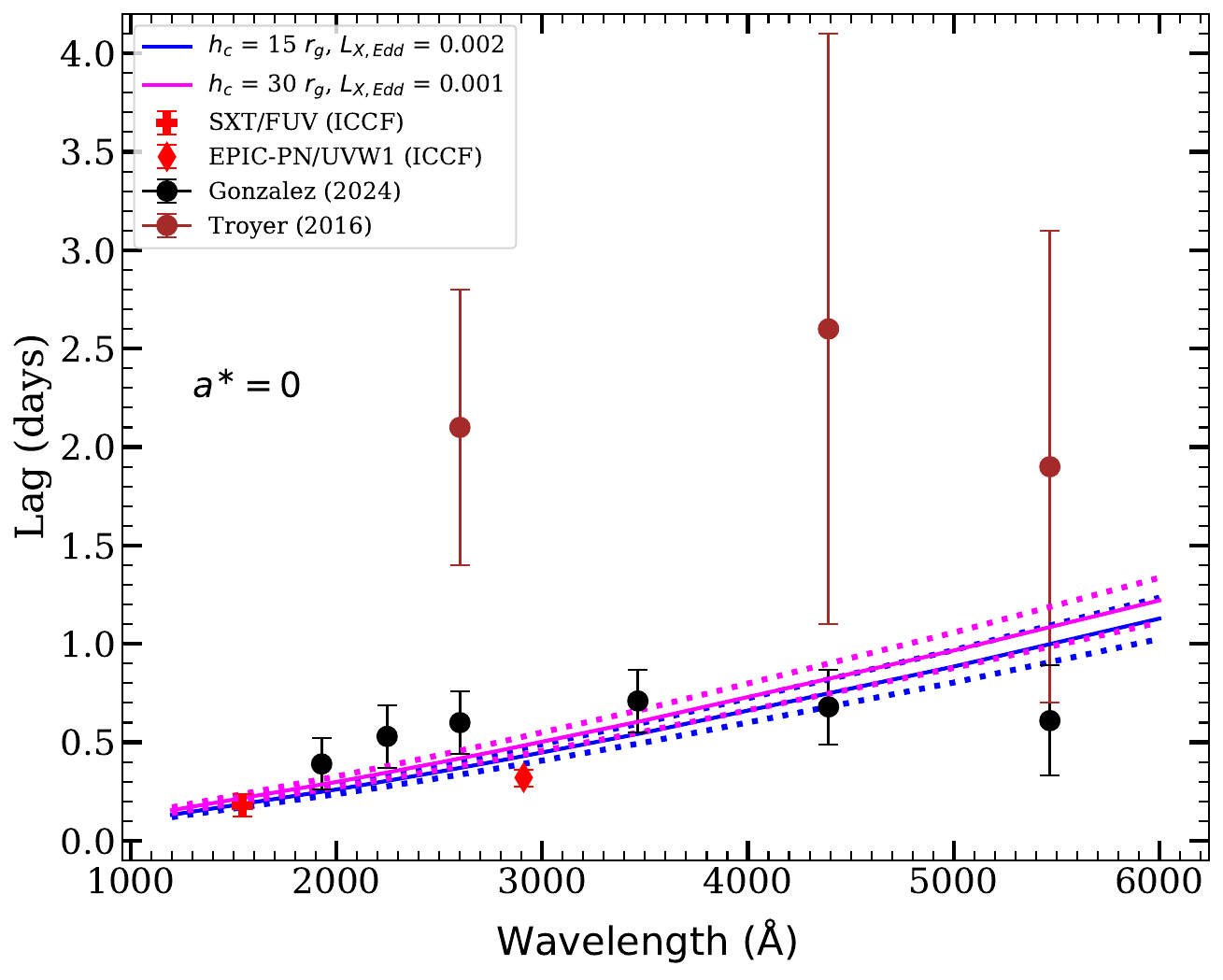}%\label{fig:lag_compare}

\caption{{\it Upper panel:} Solid lines represent the theoretical X--ray reverberation time--lags for the average corona height and unabsorbed 2-10 keV luminosity, calculated using the best--fit \textsc{kynsed} model parameters (see text for details).  
We also over-plot our SXT/FUV and EPIC--PN/UVW1 lag results using ICCF (symbols in red) and DCF (symbols in black) techniques. 
{\it Lower panel:} Same as above but the model time--lag is extrapolated to higher wavelengths including time--lags obtained by \cite{Troyer_ngc6814_2016MNRAS} and \cite{Gonzalez_2024MNRAS_ngc6814} using \swift{} observations, shown with brown and black filled circles, respectively. }
\label{fig:LAG}
\end{figure}

The blue and magenta lines in Fig. \ref{fig:LAG} show the model time--lags together with our measurements for the BH spin $a^{\ast}=0$ case. The parameter values used for computing the model lag is $M_{\rm BH} = 1.09^{+0.15}_{-0.14} \times 10^7 M_\odot $, $\dot{m}/\dot{m}_{\rm Edd}=0.1$, and $f_{\rm col}=1.7$. We used the average corona height and absorption corrected 2-10 keV luminosity ($L_{X,Edd}$) obtained from the best--fit model (see Table \ref{tab:fit_val}). The average $h_c=$ 15 and 30 $r_g$, and $L_{X,Edd}=$ 0.002 and 0.001 for \astrosat{} and XMM21 data, respectively.
The dotted lines correspond to the lags calculated using the upper and lower limits on the BH mass. We over-plot our SXT/FUV and EPIC--PN/UVW1 lag results obtained using ICCF (denoted by symbols in red) and DCF (denoted by symbols in black) methodologies (see Table \ref{tab:timelag}). The model predictions are consistent with the SXT/FUV time--lag, but they overpredict the EPIC--PN/UVW1 measurement. Perhaps this is an indication that the BH mass of NGC~6814 can be smaller than the one we used. However, it could also be the result of the fact that the XMM21 light curves are relatively short. The model predicts a time--lag between the X--rays and disc emission at  $\sim2900$~\AA\ to be less than half a day, while the XMM21 light curves only span  $\sim1.5$ days, which is only 3-4 times longer than the expected delay. Perhaps, longer light curves may be necessary to accurately detect this time--lag. For instance, \astrosat{} light curves, which can measure a time--lag of $\sim0.2$ days, extend over more than 3 days, or roughly 15 times the expected time--lag.

In the lower panel of Fig. \ref{fig:LAG}, we also compare the time--lags between the X--ray and UV/optical bands as reported by \cite{Troyer_ngc6814_2016MNRAS} and  \cite{Gonzalez_2024MNRAS_ngc6814} based on \swift{} observations from 2012 and 2022, respectively. We plot the lags with their maximum uncertainty obtained using {\it unaltered} light curves as quoted in Table~3 in \cite{Gonzalez_2024MNRAS_ngc6814}. 
Although the durations of these \swift{} observations are comparable,  the time--lags reported by \cite{Troyer_ngc6814_2016MNRAS} are larger than those measured by  \cite{Gonzalez_2024MNRAS_ngc6814} and in our study. However, the large error bars do not allow us to draw firm conclusions. It is important to note that the model time--lags depend (significantly) on the height of the X--ray corona and the X--ray luminosity of the source (as illustrated in Fig. 18 and 23 of \citealt{Kammoun_2021ApJ}). 
Both of these parameters could be significantly different during the observations used by \cite{Troyer_ngc6814_2016MNRAS} to measure time--lags and this could be the reason that the time--lag measurements are larger than the model predictions (and the measurements of \citealt{Gonzalez_2024MNRAS_ngc6814}).

Nevertheless, the overall agreement between the model time--lags and the observed time--lags is very good. We do detect an excess of the observed time--lags when compared with the model in the band between 2000--3000~\AA.
%, especially in the \cite{Troyer_ngc6814_2016MNRAS} measurement. 
This could be an indication that the diffuse emission from gas in the BLR may contribute to the inter-band lags. 
Furthermore, the observed time--lags do not indicate a strong flattening in the optical bands, when compared to the model time--lags. So, our modeling does not support the recent results claimed by \cite{Gonzalez_2024MNRAS_ngc6814}.

%%%%%%%%%%%%%%%%%%%%%%%%%%%%%%%%%%%%%%%%%%%%%%%%%%%%%%%%%%%%%%%%%%%%%%%
%------------- Figure 9: Lag vs Wavelength plot
%%%%%%%%%%%%%%%%%%%%%%%%%%%%%%%%%%%%%%%%%%%%%%%%%%%%%%%%%%%%%%%%%%%%%%%

\section{discussion and conclusions} \label{sec:discussion}

We have performed a spectral and timing analysis of NGC~6814 data from the \astrosat{}'s 2019 and  \xmm{}'s 2021 observations. The main conclusion of our work is that the X--ray disc reverberation hypothesis can explain, simultaneously, the UV/optical and X--ray variable SEDs of the source, as well as the observed time--lags,  with the same model parameters. 
We first computed the SXT/FUV and EPIC--PN/UVW1 time--lags using cross--correlation techniques and found that FUV (1541~\AA) and UVW1 (2910 \AA) band variations are well correlated ($DCF_{\rm cent} > 0.7$) with the variations in the X--ray band and are lagging by $\sim$ 15 ks and 28 ks, respectively.

Then we constructed time-resolved SEDs using appropriate X--ray spectra, collected over long periods and selected UV data points (see Fig. \ref{fig:astrosat_lc}). We applied aperture correction (only required for UVIT filter), and we subtracted the host galaxy and emission lines contributions from the UV fluxes. We note that, although we corrected our UV band fluxes for line contributions using composite quasar spectra (see Appendix \ref{appendix3}), the effect of line contributions to the UV continuum fluxes may be different during our observations and it varies over the years. For that reason, we also fitted the data without subtracting the line flux contribution to that data, and we found that the best--fit results do not change significantly (values are quoted in Table \ref{tab:fit_val}).

We then performed simultaneous spectral fitting of four broadband SEDs (3 from \astrosat{} and 1 from \xmm{}'s 2021 observations) using the \textsc{kynsed} model. The model fits the variable SEDs well, indicating that X--ray reverberation of the accretion disc in NGC~6814 can explain the observed X--ray and UV variations of the source, assuming an inclination of $\theta=70^\circ$, and an $f_{\rm col}$ of $1.7$. The observed variability can be explained by assuming a constant accretion rate of 
$\dot{m}/\dot{m}_{\rm Edd}\sim 0.1$, i.e. the observed variations are merely due to variations of $L_{\rm transf}/L_{\rm disc}$ and/or variations of the X--ray source height.  
The best--fit values indicate a moderately spinning BH i.e. $a^{\ast}=0-0.5$ and corona height ranging between $\sim7.5-35~r_g$. 
We also found that $R_{\rm out}$ does not affect our best--fit results.

We found that both assumptions of an X--ray corona which is powered by the accretion disc, as well as a corona powered by an external source (i,.e. the case of positive and negative $L_{\rm transf}$) provide a good fit to the SEDs and yield similar best--fit parameter values. 
In the case when the accretion process powers the X--ray corona, our best--fit results indicate that 10--20\% of accretion power below $R_{t}$ may get transferred to the corona. We obtained the color correction factor, $f_{\rm col}=1.7$ which is consistent with the moderate variation of $f_{\rm col}$ between 1.4--2 as suggested by \cite{Davis_and_ElAbd_2019ApJ}. They have also provided an analytic expression for $f_{\rm col}$ as a function of $M_{\rm BH}$ and $\dot{m}/\dot{m}_{\rm Edd}$ for $a^\ast=0$ (see Eq. 10 in their paper). Using this equation, we calculated $f_{\rm col}\approx1.6$ (for $\alpha=0.1$) which is very close to the 1.7 we used for fitting SEDs.

The obtained coronal heights from our best--fit result ($7.5-35~r_g$) are quite consistent with the heights found in other AGNs (e.g. \citealt{Kammoun_2021MNRAS}). The estimated size of the corona ranges between $\sim 6-10~r_g$ which is similar to what has been found in recent studies by \cite{Gallo_ngc6814_2021ApJ} and \cite{Kang_2023MNRAS}.
The changing height and $R_c$ in different spectra are possible due to dynamical variability in the height and size of the corona (\citealt{Panagiotou2022ApJ, Zhang_2023MNRAS}).

We utilized the best--fit parameter values listed in Table \ref{tab:fit_val} for modelling time--lag. We have used the updated analytic equations for computing time--lag as a function of wavelength (shown in Fig. \ref{fig:LAG}).
While the model aligns well with the SXT/FUV time--lag, it overestimates the EPIC--PN/UVW1 lag, possibly due to the shorter duration of the XMM21 light curves or an overestimated black hole mass. Comparisons with previous studies by \cite{Troyer_ngc6814_2016MNRAS} and \cite{Gonzalez_2024MNRAS_ngc6814} reveal larger time--lags in \cite{Troyer_ngc6814_2016MNRAS} data, which could be due to different X--ray luminosity and/or corona height during their observation. With the slight excess in the 2000--3000~\AA~ range, the overall agreement between model predictions and observations is strong, with indications that diffuse emission from the broad-line region may contribute to the observed lags.

Our analysis establishes the usefulness of \textsc{kynsed} model for spectral--timing analysis for a lamp--post geometry of corona. We did not find any evidence of an extremely truncated or non--standard disc geometry in NGC 6814, as observed by \cite{Gonzalez_2024MNRAS_ngc6814} in the original paper. Future studies with longer observations and improved sampling may yield further insights into the geometry and physical conditions near the black hole in this source.\\

%\vspace{2 em}
%\begin{acknowledgments}
\section*{Acknowledgements}
 
We thank the anonymous reviewer for their insightful and valuable feedback, which helped enhance the quality of this manuscript. This study utilized data from the \astrosat{} mission, conducted by the Indian Space Research Organisation (ISRO) and archived at the Indian Space Science Data Centre (ISSDC). Additionally, the research benefited from data obtained through the Soft X--ray Telescope (SXT), developed at TIFR, Mumbai. The UVIT project was a collaborative effort involving IIA Bengaluru, IUCAA Pune, TIFR Mumbai, and various ISRO and CSA centres. The SXT Point of Contact (POC) at TIFR and the UVIT POC at IIA Bengaluru played essential roles in validating and releasing the data via the ISSDC archive. The UVIT data were processed using the CCDLAB pipeline. This study also utilized archived \xmm{} observations. 
This research utilized the \textsc{xspec} and \textsc{sherpa} software packages, as well as \textsc{python} and \textsc{julia} programming languages. Additionally, it made use of the SIMBAD and NED databases.
The author thanks Prof. Misty. C. Bentz for providing the observed host galaxy starlight flux density at 5100~\AA~ through personal communication.
The author would also like to thank Shrabani Kumar and Piyali Ganguly for their valuable discussions.

%\end{acknowledgments}

\vspace{5mm}
%\facilities{HST(STIS), Swift(XRT and UVOT), AAVSO, CTIO:1.3m,



\appendix \label{appendix}

%%%%%%%%%%%5----------Appendix 1   ----------%%%%%%%%%%%%%%%%%%

\section{UVIT/F154W Aperture correction }\label{appendix1}
\restartappendixnumbering

%%%%%%%%%%%%%%%%%%%%%%%%%%%%%%%%%%%%%%%%%%%%%%%%%%%%%%%%%%%%%%%%%%%%%%%%%
%---------------Figure A1: radial profile images-------------------
%%%%%%%%%%%%%%%%%%%%%%%%%%%%%%%%%%%%%%%%%%%%%%%%%%%%%%%%%%%%%%%%%%%%%%%%%

\begin{figure}[!ht]
\centering
\includegraphics[width=0.45\columnwidth]{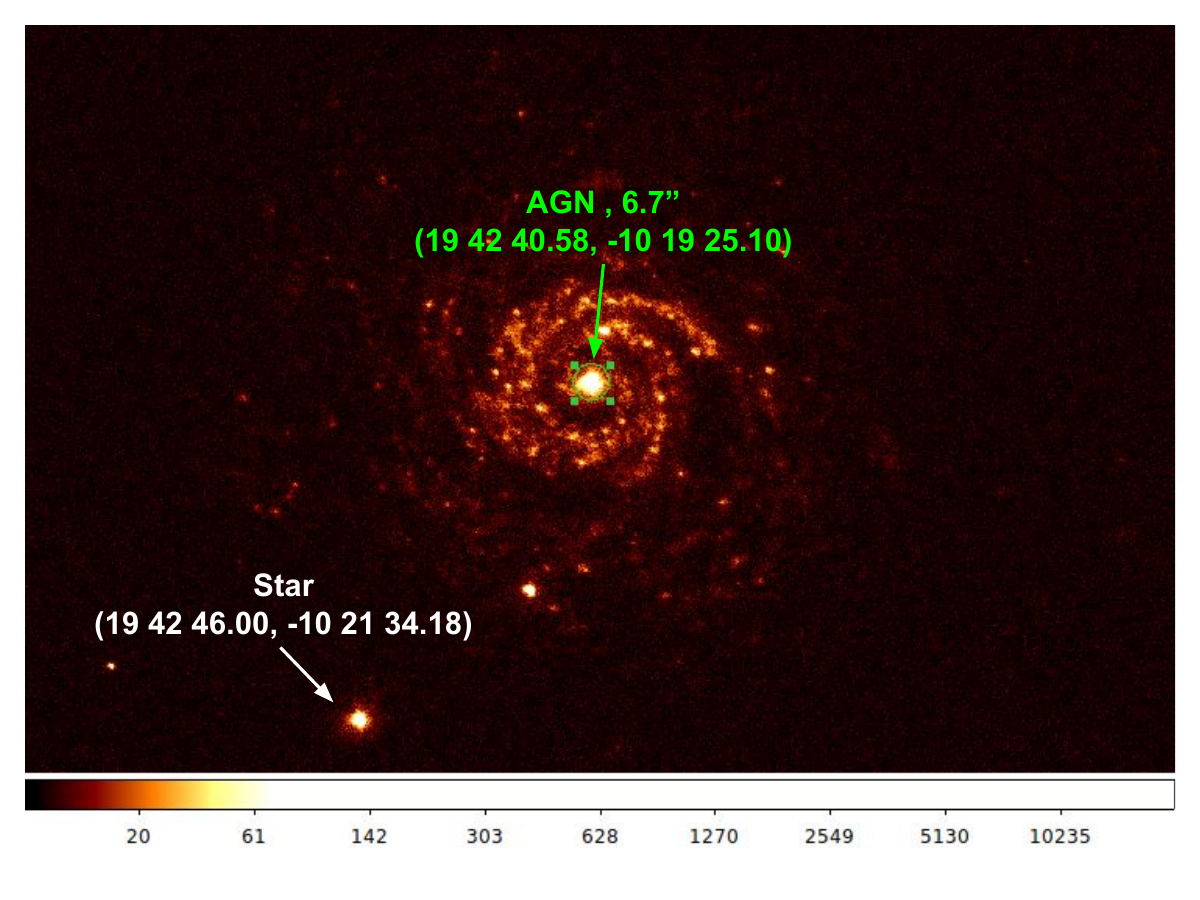}\label{fig:uvit_source_image}
%\hfill
%
\includegraphics[width=0.45\columnwidth]{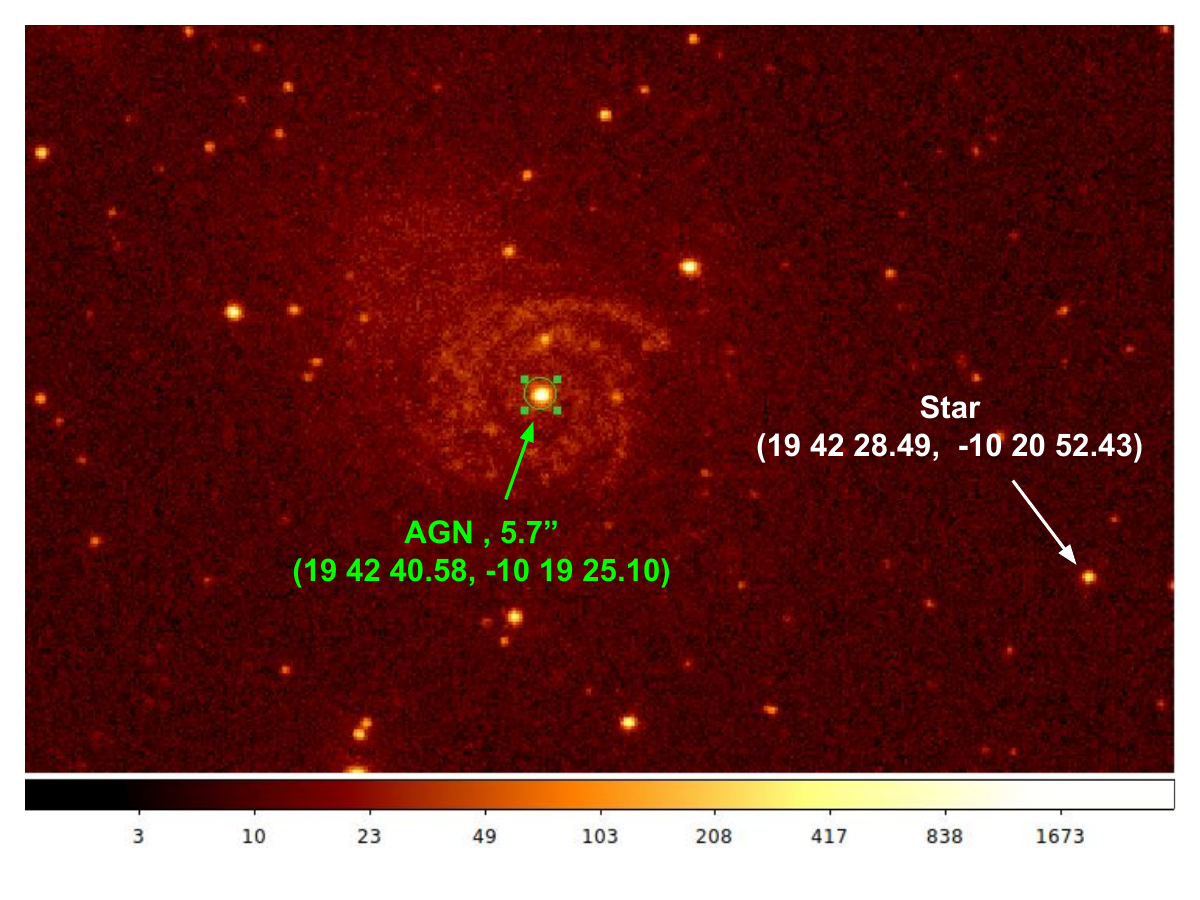}\label{fig:om_source_image}
\caption{{\it Left panel:} Image of NGC~6814 observed through the UVIT/F154W filter. {\it Right panel:} Image of NGC~6814 observed through OM/UVW1 filter. The source extraction regions for the AGN (highlighted in green) and the stars used for PSF estimation via radial profile analysis are indicated. }
\label{fig:source_image}
\end{figure}

%%%%%%%%%%%%%%%%%%%%%%%%%%%%%%%%%%%%%%%%%%%%%%%%%%%%%%%%%%%%%%%%%%%%%%%%%
%---------------Figure A2: radial profile images-------------------
%%%%%%%%%%%%%%%%%%%%%%%%%%%%%%%%%%%%%%%%%%%%%%%%%%%%%%%%%%%%%%%%%%%%%%%%%

\begin{figure}[!ht]
\centering
\gridline{\fig{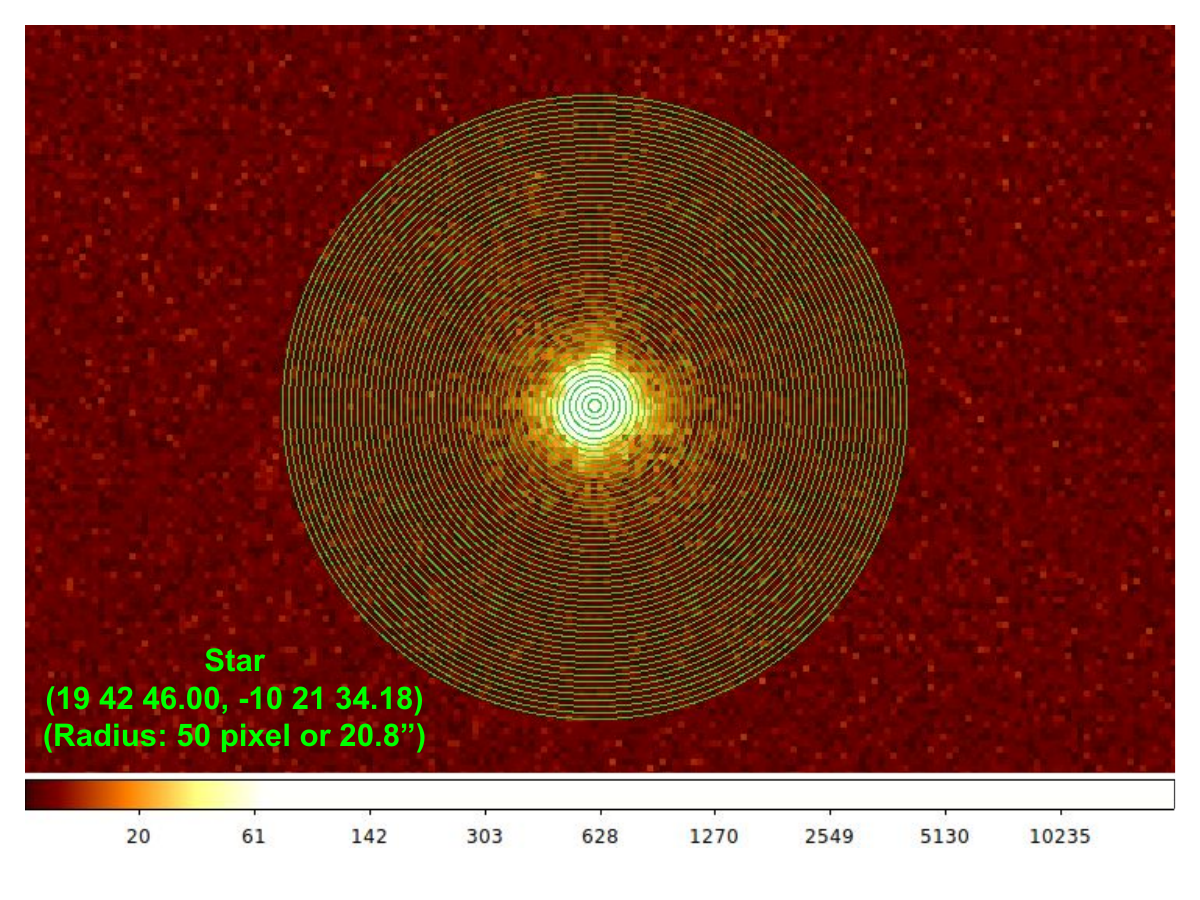}{0.45\textwidth}{}
          \fig{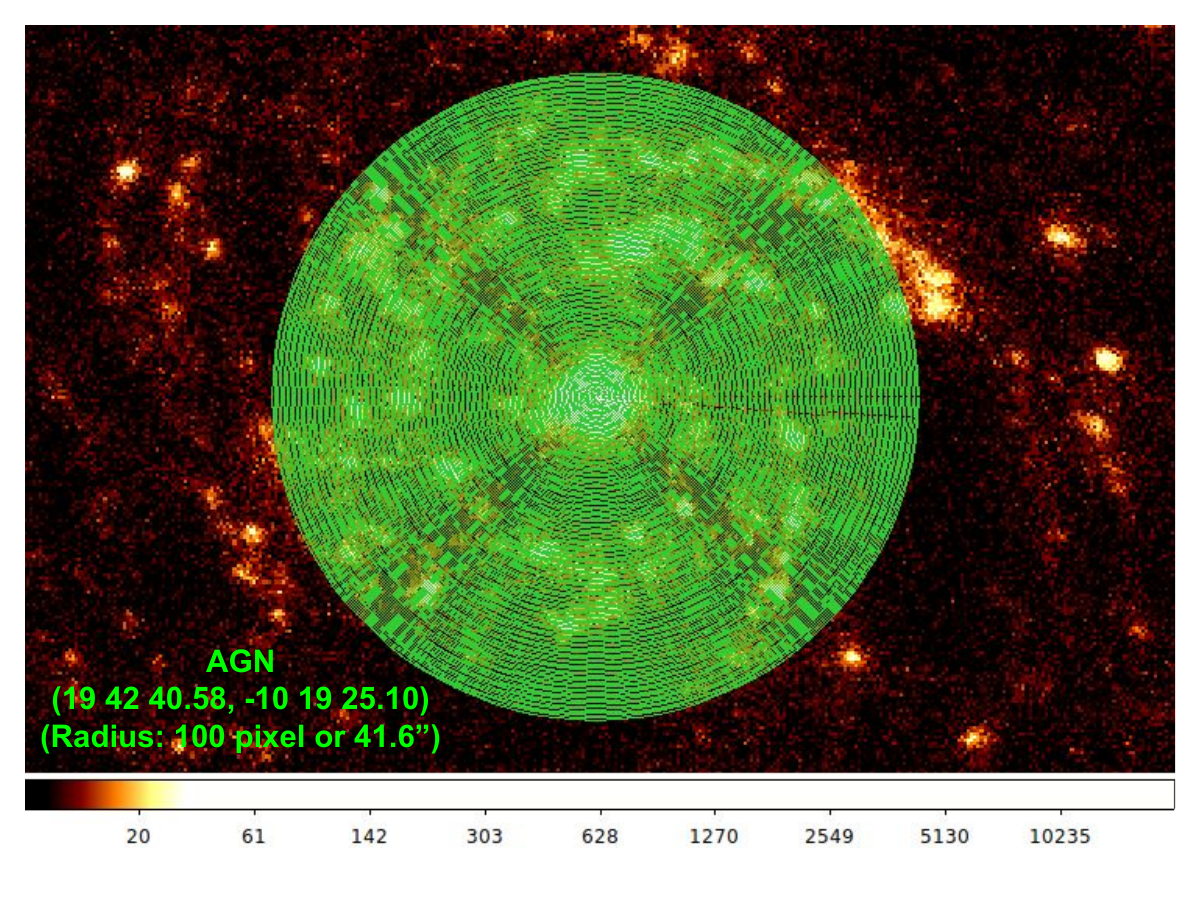}{0.45\textwidth}{}}
\gridline{\fig{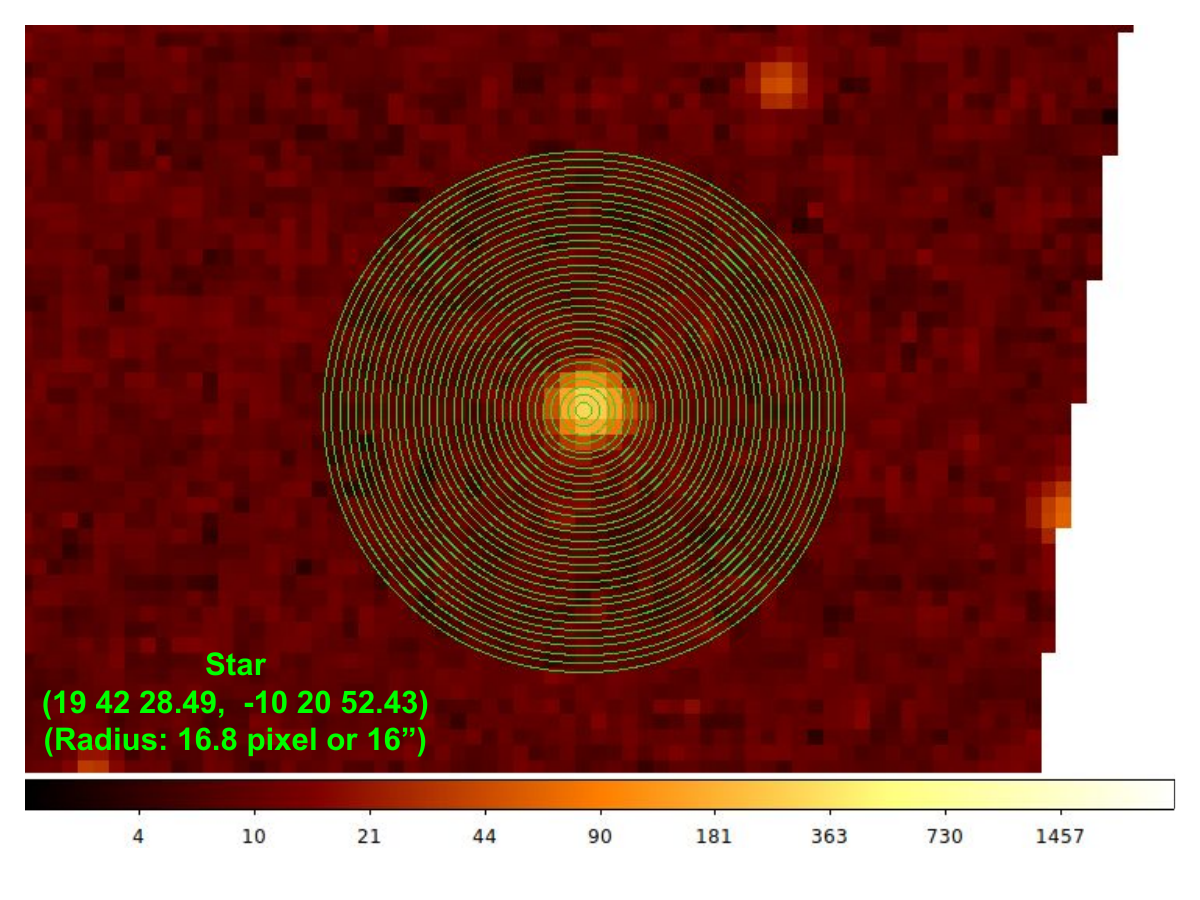}{0.45\textwidth}{}
          \fig{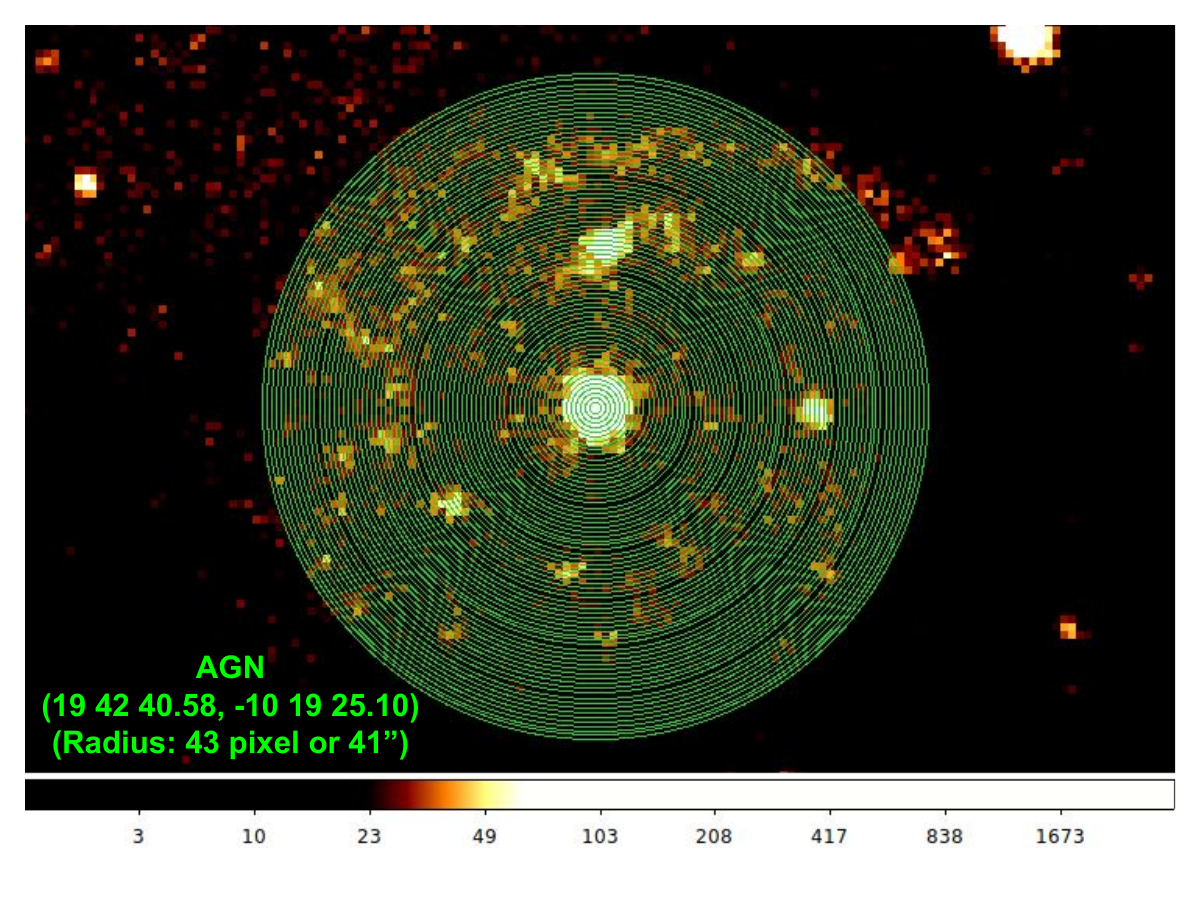}{0.45\textwidth}{}}
\caption{ {\it Left panels:}  Circular annuli centred on the star in the UVIT/F154W and OM/UVW1 images (upper left and lower left panels, respectively), used to obtain the radial profile from DS9 software.   {\it Right panels:} Same for AGN. The extraction radius is mentioned in the left corner of the images.}
\label{fig:RF_images}
\end{figure}

%%%%%%%%%%%%%%%%%%%%%%%%%%%%%%%%%%%%%%%%%%%%%%%%%%%%%%%%%%%%%%%%%%%%%%%%%
%---------------Figure A3: radial profile images-------------------
%%%%%%%%%%%%%%%%%%%%%%%%%%%%%%%%%%%%%%%%%%%%%%%%%%%%%%%%%%%%%%%%%%%%%%%%%

\begin{figure}[!ht]
\centering
\includegraphics[width=0.5\columnwidth]{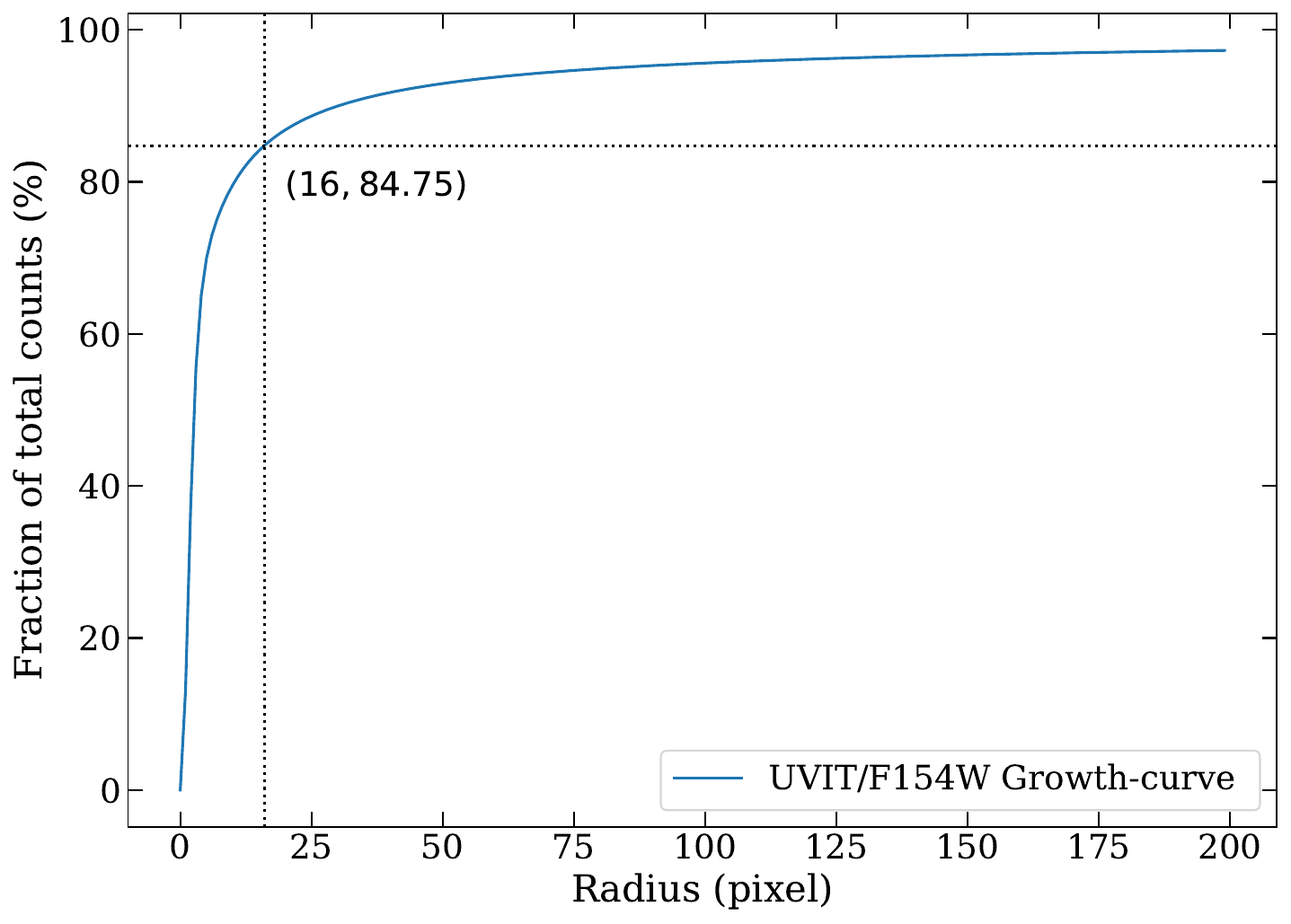}
\caption{The PSF growth curve for the UVIT/F154W filter was obtained for a star in the field of view (FOV). The y-axis shows the fraction of total counts (in percentage) enclosed within a circle of the radius indicated on the x-axis. The intersection point of the vertical and horizontal lines represents the radius of the source extraction region and the percentage of total counts enclosed within it.}
\label{fig:psf_growth}
\end{figure}

The point spread function (PSF) of the UVIT filters is characterized by a narrow core and an extended pedestal. This pedestal results from scattering due to mechanical obstructions and the surface roughness of mirrors and filters (\citealt{Tandon_2020AJ}). The fractional flux contribution of the pedestal can be significant, especially when the source aperture size is small, and it varies across different filters. To quantify the flux missing outside a 16~pixels radius (the source extraction region), we analyzed the radial profile of a bright star in the field of view (FOV), GPM 295.691591-10.359400 (RA: 19 42 46.001, Dec: -10 21 34.180), marked in the left panel of Fig. \ref{fig:source_image}. 
We derived the growth curve of the PSF by fitting the radial profile from the co-added average UVIT/F154W image (see section \ref{subsec:astrosat_data}). We obtained the radial profile using DS9 software \citep{DS9_2000ascl}. The extraction region to obtain the radial profile is shown in the upper left panel of Fig. \ref{fig:RF_images}.

We have used  {\sc sherpa} software \citep{freeman_2011ascl_sherpa} for fitting the radial profiles. The core and wing components of the radial profile were modelled using double Moffat functions, described by, $M(x) = A_m \left[ 1+ \left(\frac{x}{\alpha}\right)^2 \right]^{-\beta}$  $ \left(FWHM =2 \alpha \sqrt{2^{\frac{1}{\beta}}-1}\right)$\footnote{FWHM: Full Width at Half Maximum}, plus a constant background obtained from the source free region. The FWHM for the best--fit model profile of the PSF is 1.5 arcsec. We then integrated the double Moffat functions from zero to infinity to obtain the growth curve, as illustrated in Fig. \ref{fig:psf_growth}. The growth curve analysis revealed that approximately 15\% of the total flux is located outside the 16~pixels radius, indicating a significant flux loss. Consequently, we corrected this by adding the missing 15\% to the \astrosat{}'s FUV fluxes before SED fitting.

%%%%%%%%%%%5----------Appendix 2   ----------%%%%%%%%%%%%%%%%%%

\section{Host Galaxy contribution}\label{appendix2}
\restartappendixnumbering

\begin{figure}[!ht]
\centering
\includegraphics[width=0.48\columnwidth]{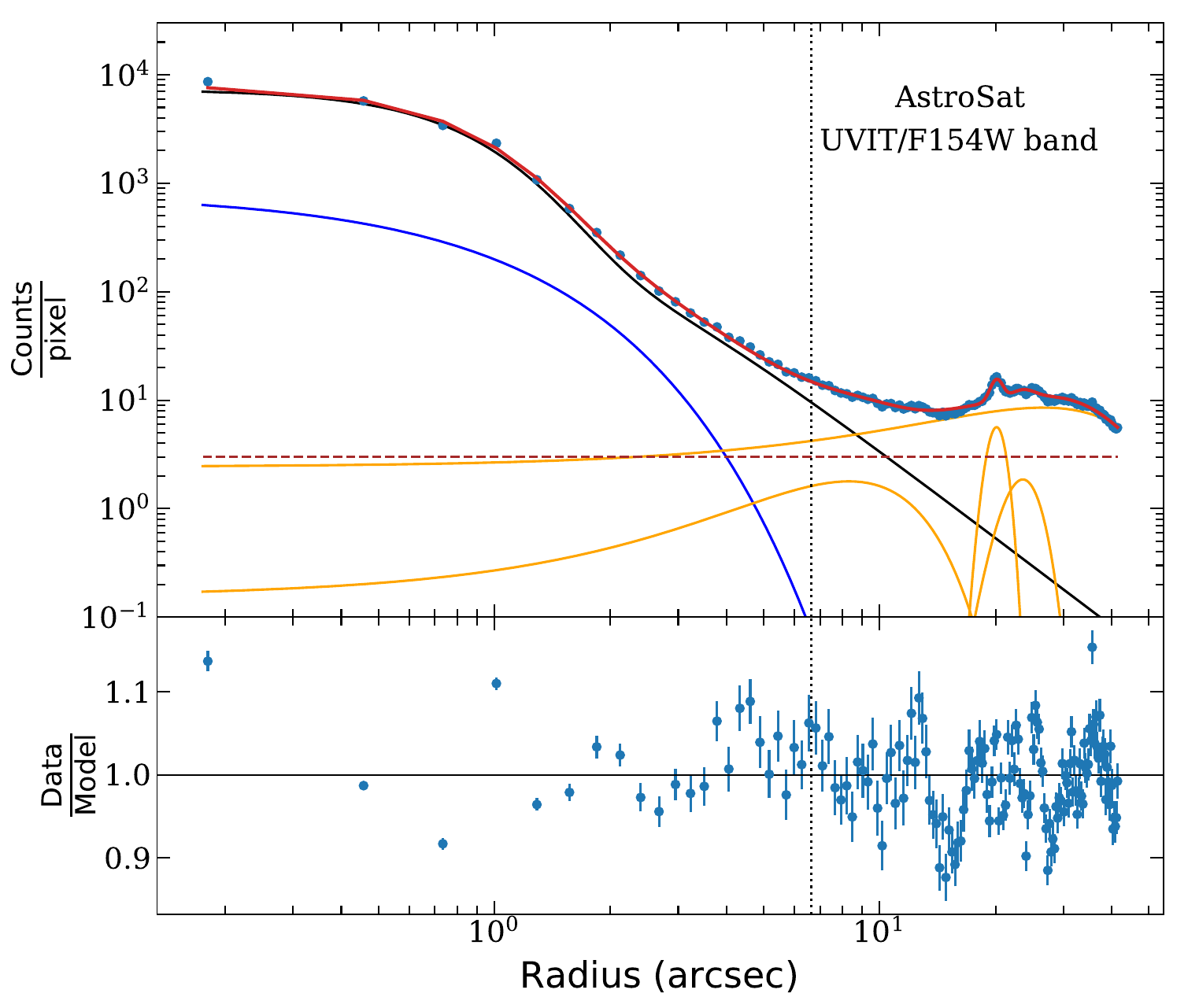}
%\hfill
%
\includegraphics[width=0.48\columnwidth]{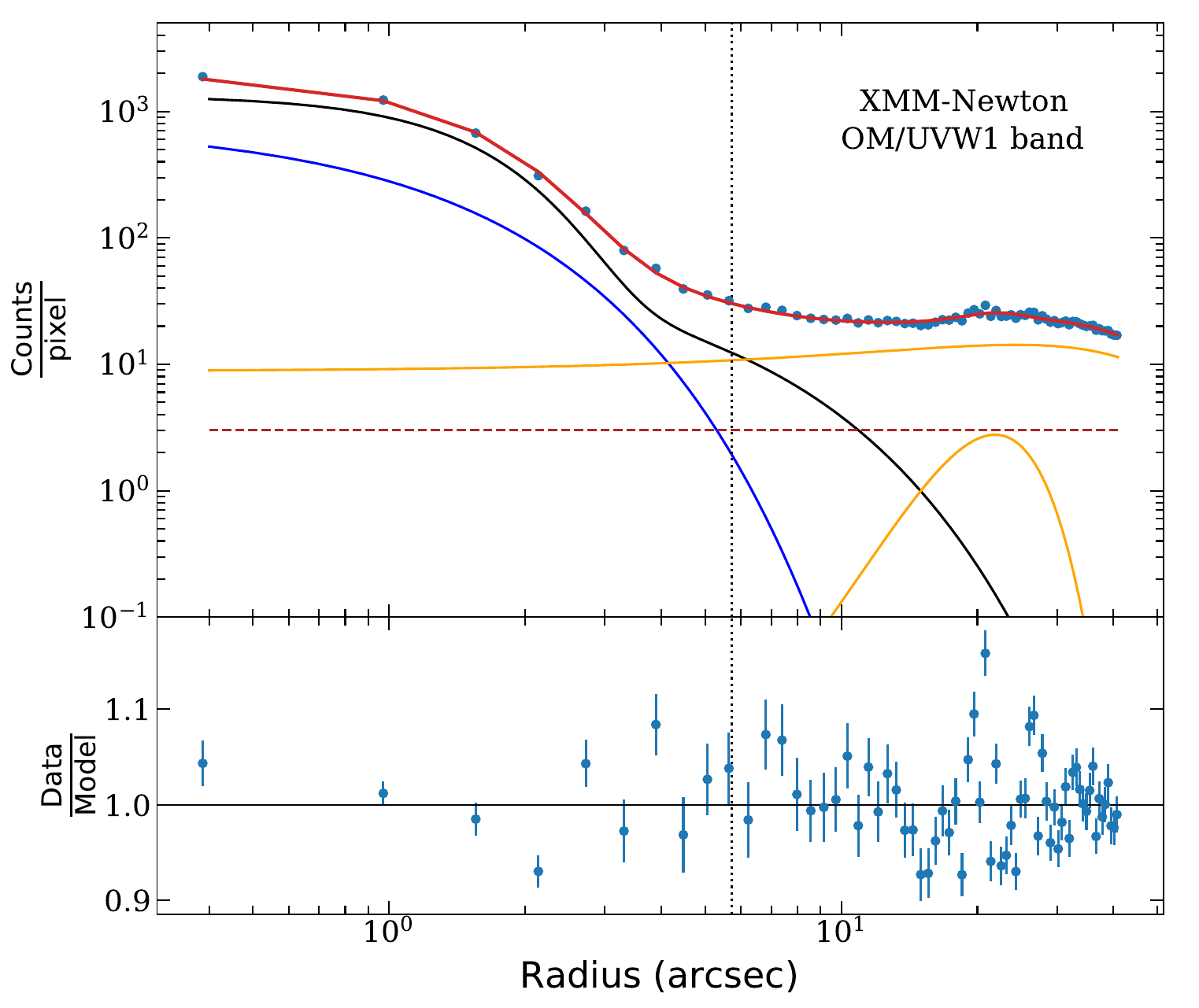}
\caption{Radial profile fits of NGC~6814 in UVIT/F154W band (left figure) and OM/UVW1 band (right figure). The radial profile data points are indicated by blue dots.
The profiles are fitted with the PSF of the respective instruments to model the AGN emission (depicted in black) and an exponential function to represent the host galaxy's emission (depicted in blue). Additionally, multiple Gaussian functions (shown in orange) are used to model the structures detected beyond 10 arcsec, likely corresponding to star-forming regions in the galaxy's spiral arms. The dotted black vertical lines represent the source extraction region used in the analysis and for estimating the host galaxy contribution. The lower panels of both figures display the residuals of the fits.
 }
\label{fig:radial_profile}
\end{figure}

The diffuse emission from the host galaxy often contaminates the observed UV/optical fluxes from the broadband images. To estimate the host galaxy contribution in the UVIT/F154W filter band, we performed a radial profile analysis of NGC~6814 using the merged average UV image (see section \ref{subsec:astrosat_data}).  We extracted the radial profile of the AGN source from 100~pixels ($\sim 42$ arcsec) region using DS9 software (upper right panel in Fig. \ref{fig:RF_images}).
We used the best--fit PSF profile of the FOV star (explained in Appendix \ref{appendix1}) with the variable amplitude of the Moffat Functions to fit the radial profile of the central part of NGC~6814, which should represent the AGN flux profile.  For the host galaxy emission, we used an exponential function, $I(x) = I_0\times exp^{\left(\frac{-x}{d}\right)}$ plus a constant for the background. We also needed to consider multiple Gaussian functions $\left(f(x) = A_G \times exp\left[ \frac{-4log(2)(x-x_0)^2}{fwhm^2}\right] \right)$ to account for emission features beyond 10~arcsec (see the left panel in Fig.~\ref{fig:radial_profile}). Their amplitude is small when compared to the amplitude of the AGN, however, their presence is required to improve the fit. These emission features are most likely due to the star-forming regions in the galaxy's spiral arms that we can see in the images (right panels in Fig. \ref{fig:RF_images}). 
From the best--fit model,  we calculated the host galaxy contribution to be $\sim 12.2$\% within the 16~pixels (6.7 arcsec) radius (this is the size of the aperture we used to extract the light curve from the central region in this source).

For the OM/UVW1 filter, we utilized one of the last two orbital images to estimate the host galaxy's contribution (as the flux from these orbits is employed for spectral fitting). We used the well-isolated and bright star, GPM~295.618808-10.347919,  (RA: 19 42 28.49, Dec: -10 20 52.43)  to estimate the instrument's PSF (marked in the right panel of Fig. \ref{fig:source_image}). 
We extracted the radial profile of the AGN source from a circular region of 41~arcsec radius, similar to what we used for UVIT. The extraction regions to obtain the radial profiles are shown in the lower panels in Fig. \ref{fig:RF_images}.
 We fitted the core of the PSF with a  Moffat function and the extended wings with the exponential function, and a constant background. The FWHM of the PSF is 2.7 arcsec.
We employed a similar approach to fit the radial profile of the AGN source as described for UVIT. The best--fit plot is shown in the right panel in Fig. \ref{fig:radial_profile}.
The host galaxy contribution is found to be $\sim29.5$~\% within a 6~pixels (5.7 arcsec) radius. Before spectral fitting, we subtracted the host galaxy contribution from the respective UV fluxes.

To check the validity of our estimates for the host galaxy contribution, we have used Sc and Sb spiral galaxy templates from SWIRE template library\footnote{\url{http://www.iasf-milano.inaf.it/~polletta/templates/swire_templates.html}} \citep{Polletta_2007ApJ}. NGC~6814 is classified as an SAB(rs)bc galaxy \citep{Vaucouleurs_1991rc3D}, so we expect that our calculated host galaxy flux should match with either of the two templates.  These templates were scaled according to the host galaxy flux density of NGC~6814 at $\rm 5100$\AA, measured within a 5~arcsec radius aperture. The observed host galaxy flux density, $\rm F_{5100} = 6.27 \pm 0.31 ~ergs~s^{-1}cm^{-2}\text{\AA}^{-1}$, was provided by M. Bentz (personal communication), based on the detailed work on the NGC~6814 host galaxy morphology \cite{Bentz_2013ApJ}. 

\begin{figure}[!ht]
\centering
\includegraphics[width=0.5\columnwidth]{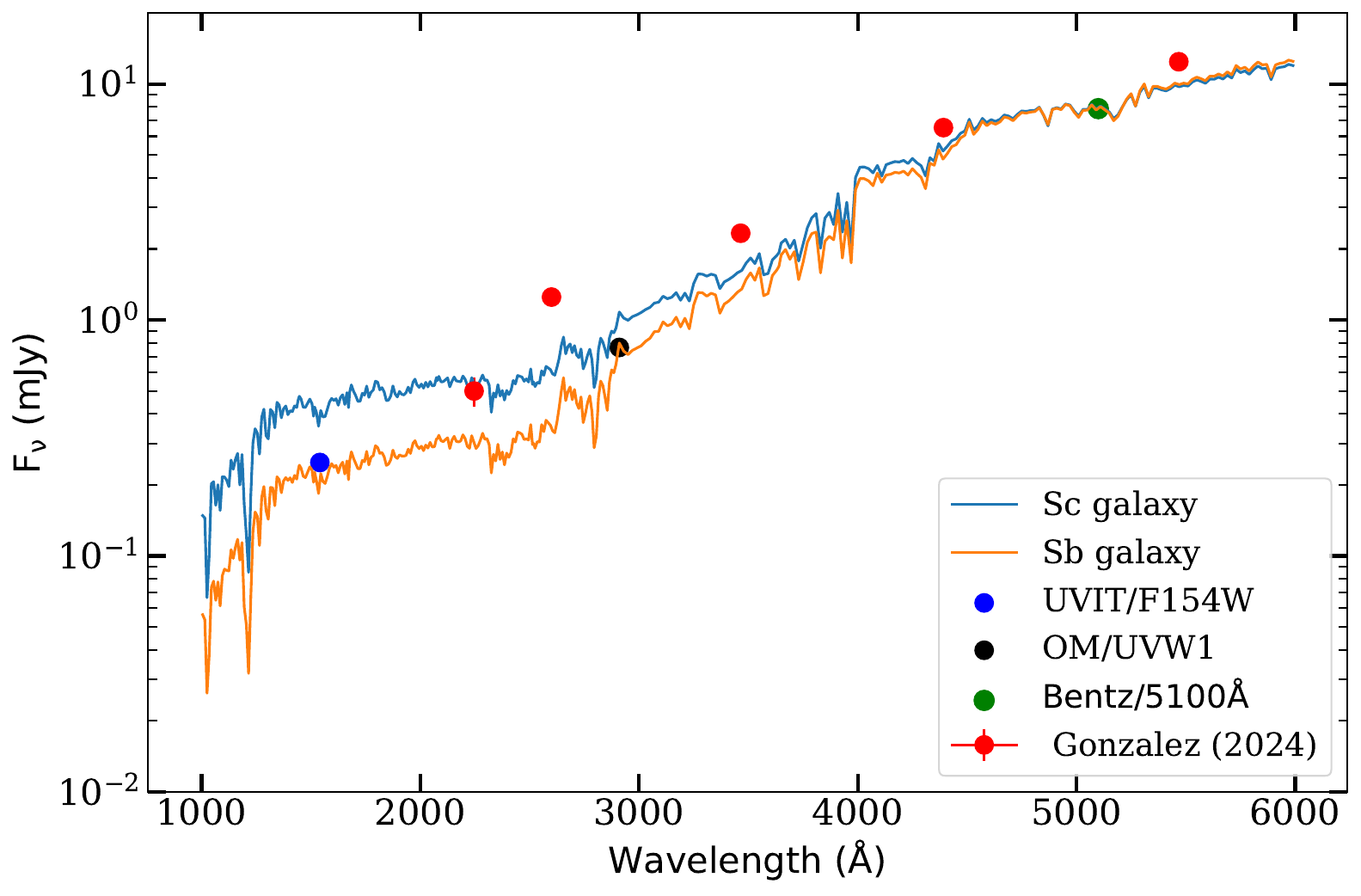}
\caption{The blue and orange solid lines represent Sc and Sb host galaxy templates, respectively. These templates were scaled using the host galaxy flux density of  NGC~6814 at 5100~\AA~within 5 arcsec radius aperture,  $\rm F_{5100} = 6.27 \pm 0.31 ~ergs~s^{-1}cm^{-2}\text{\AA}^{-1}$ (courtesy: Prof. Misty C. Bentz for providing this value based on her work in \citealt{Bentz_2013ApJ}). For scaling purposes, we corrected the given $\rm F_{5100}$ for galactic extinction using V-band extinction $\rm A_V =0.38$ (depicted in green). The filled circles in blue and black are the host galaxy estimates in UVIT/F154W and OM/UVW1 filter bands, respectively, as obtained from radial profile analysis (explained in Appendix \ref{appendix2}). Additionally, the host galaxy contribution calculated by \cite{Gonzalez_2024MNRAS_ngc6814} using flux-flux analysis is shown in red.}
\label{fig:host_galaxy_template}
\end{figure}

For the purpose of accurate scaling, we corrected this $F_{5100}$ value for galactic extinction, utilizing a V-band extinction coefficient of  $A_V =0.38$. The resulting galactic flux is shown by the green circle in Fig. \ref{fig:host_galaxy_template}. We over-plot our estimates of the galactic flux within 5~arcsec radius aperture in UVIT/F154W and OM/UVW1 bands, obtained through the radial profile analysis we discussed above, as shown by the blue and black filled circles in Fig. \ref{fig:host_galaxy_template}, respectively.  The starlight contribution from the host galaxy was calculated to be $\rm \sim 0.25 ~mJy$ and $\rm \sim 0.76~ mJy$
in the UVIT/F154W and OM/UVW1 bands, respectively. These values are corrected for the galactic extinction using the prescription given by \cite{Cardelli_1989ApJ_redden}. Our values align well with the Sb galaxy template.

Filled red circles in Fig. \ref{fig:host_galaxy_template}, indicate the host galaxy flux calculated by \cite{Gonzalez_2024MNRAS_ngc6814} within the same aperture,  using the so-called ``flux-flux analysis" method. 
Their values are larger than ours, and in fact, they are larger than even the Sc galaxy template. The ``flux-flux" analysis overestimates the host galaxy flux in other AGNs as well. For example, flux-flux analysis in the case of NGC~5548 also overestimates the host galaxy flux compared to the image decomposition method (for reference, see Table 6 in \citealt{Starkey_2017ApJ_ngc5548} and Table 5 in \citealt{fausnaugh_2016ApJ_ngc5548}).
This comparison highlights the consistency and differences between various methodologies for determining the host galaxy's starlight contribution to the observed UV fluxes.

%%%%%%%%%%%%%%%%%%%%%%%%%%%%%%%%%%%%%%%%%%%%%%%%%%%%%%%%%%%%%%%%%%%%%%%%%
%---------------Figure B1: radial profile -------------------
%%%%%%%%%%%%%%%%%%%%%%%%%%%%%%%%%%%%%%%%%%%%%%%%%%%%%%%%%%%%%%%%%%%%%%%%%

%%%%%%%%%%%%%%%%%%%%%%%%%%%%%%%%%%%%%%%%%%%%%%%%%%%%%%%%%%%%%%%%%%%%%%%%%
%---------------Figure B2: host galaxy flux-------------------
%%%%%%%%%%%%%%%%%%%%%%%%%%%%%%%%%%%%%%%%%%%%%%%%%%%%%%%%%%%%%%%%%%%%%%%%%

%%%%%%%%%%%5----------Appendix 3   ----------%%%%%%%%%%%%%%%%%%

\section{Emission line contribution}\label{appendix3}
\restartappendixnumbering

Emission lines originating from the narrow-line and broad-line regions, blended Fe~II emission and Balmer continuum can substantially impact the flux observed through broadband filters in the optical/UV spectrum.  In our observations, the broad emission line of C IV$\lambda1549$ predominantly contributes to the flux within the UVIT/F154W filter bandpass. At the same time, the OM/UVW1 band is affected by the Mg~II$\lambda2798$ emission line, the Balmer continuum and Fe~II emission. 

\begin{figure}[!ht]
\centering
\includegraphics[width=0.5\columnwidth]{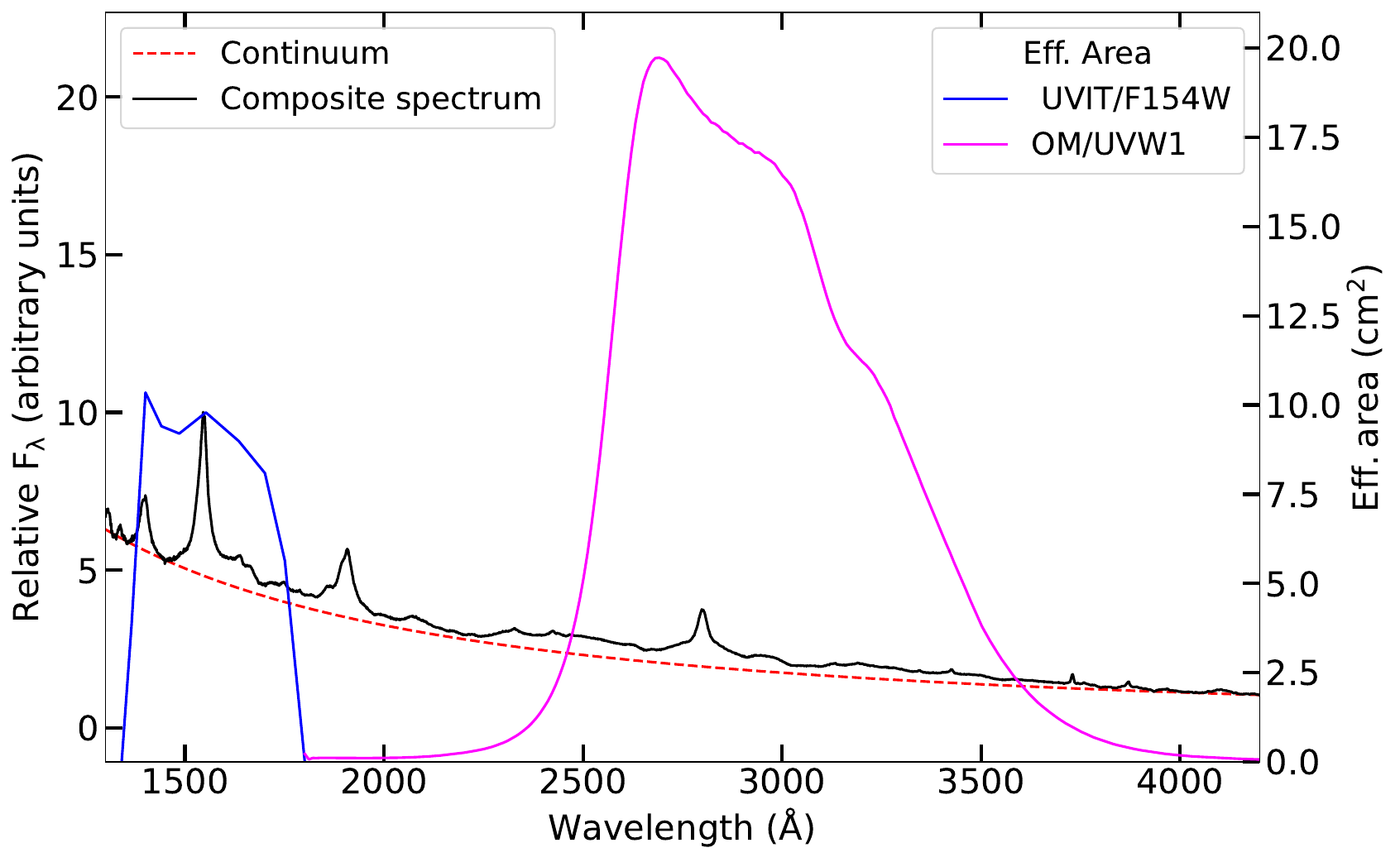}
\caption{The composite quasar spectrum of \cite{vanderberk2001AJ} and continuum fitted with power-law are shown in solid black and dashed red lines, respectively. We also over-plot the effective areas of broadband filters UVIT/F154W   and OM/UVW1, used to estimate the emission line contributions above the continuum.}
\label{fig:composite_spectra}
\end{figure}

To estimate the emission line contribution in the measured UV fluxes, we have used the composite quasar spectra given by \cite{vanderberk2001AJ} and followed the same procedure as in \cite{gulab2021MNRAS}. The composite spectrum, continuum fitted with power-law, and the effective areas of broadband filters UVIT/F154W \citep{Tandon_2020AJ} and OM/UVW1\footnote{\url{https://www.cosmos.esa.int/web/xmm-newton/om-response-files}} are shown in Fig. \ref{fig:composite_spectra}.
We obtained $\sim16.7\%$ and $\sim21\%$ emission line contribution to the intrinsic continuum flux in UVIT/F154W and OM/UVW1 filter bands, respectively. We subtracted these contributions from respective UV fluxes before spectral analysis.

%%%%%%%%%%%%%%%%%%%%%%%%%%%%%%%%%%%%%%%%%%%%%%%%%%%%%%%%%%%%%%%%%%%%%%%%%
%---------------Figure C1: composite spectra-------------------
%%%%%%%%%%%%%%%%%%%%%%%%%%%%%%%%%%%%%%%%%%%%%%%%%%%%%%%%%%%%%%%%%%%%%%%%%

\newpage

\bibliography{main}{}
\bibliographystyle{aasjournal}



\end{document}